\documentclass[10pt]{article}

\usepackage[a4paper,left=22mm,right=22mm,
            top=22mm,bottom=22mm]{geometry}

\usepackage{amsmath,amssymb}
\usepackage{amsthm}
\usepackage{mathrsfs}

\usepackage{graphicx}
\usepackage{caption}
\usepackage{dblfloatfix}

\usepackage{booktabs}
\usepackage{multirow}
\usepackage{tabularx}
\usepackage{array}
\usepackage{ragged2e}

\usepackage[table]{xcolor}

\definecolor{octamode}{RGB}{235,245,255}
\definecolor{classA}{gray}{0.92}

\usepackage[version=4]{mhchem}
\usepackage{chemformula}

\usepackage{algorithm}
\usepackage{algorithmicx}
\usepackage{algpseudocode}
\usepackage{listings}

\usepackage{textcomp}
\usepackage{manyfoot}
\usepackage[title]{appendix}
\usepackage{multicol}
\usepackage{csquotes}
\usepackage{authblk}

\usepackage[numbers,sort&compress]{natbib}
\usepackage{xurl}
\usepackage[hidelinks,breaklinks=true]{hyperref}

\newcommand{\val}[1]{#1}

\title{Phonon evidence for a partial, re-entrant spin reorientation near room temperature in LaFe$_{1-x}$Ni$_x$O$_3$}

\author[1]{Subha Maity}
\author[1]{Dushyant Singh}
\author[1]{Vigneshwari SR}
\author[1]{Tharundev VV}
\author[1]{Navneet Sharma}
\author[1]{Subhavenkadesh SS}
\author[1]{Krista R. Khiangte\textsuperscript{*}}

\affil[1]{Department of Physics, Indian Institute of Technology Gandhinagar,
Gandhinagar 382355, Gujarat, India}

\affil[*]{Corresponding author: krista.khiangte@iitgn.ac.in}

\date{}

\begin{document}

\maketitle

\begin{abstract}
Spin reorientation near room temperature offers a route to controlling antiferromagnetic order, yet LaFeO$_3$, which lacks a magnetic rare-earth ion, is generally expected to retain its spin orientation along the $a$ axis below $T_{\mathrm N}\approx740$~K. Here, we investigate LaFe$_{1-x}$Ni$_x$O$_3$ ($x=0$ and 0.1) using temperature-dependent magnetization and Raman spectroscopy over 200--400~K. Both compositions exhibit magnetic anomalies at 285.6 and 323~K, respectively, accompanied by correlated deviations of 12 Raman modes from their conventional anharmonic behavior. Principal component analysis and a global fitting approach reveal a common temperature-dependent phonon response, with mode-dependent amplitudes and signs. To interpret these anomalies, we develop a Landau-based spin-phonon coupling model describing complete, partial re-entrant, and incipient spin reorientation. Comparison with the experimental Raman profiles favors a partial, re-entrant rotation of the N\'eel vector in both compositions, centered near 290~K. Ni substitution narrows the characteristic reorientation interval from approximately 116 to 84~K and enhances the magnetization anomaly by a factor of four. An additional feature near 262~K in LaFeO$_3$ is consistent with the Morin transition of trace $\alpha$-Fe$_2$O$_3$. These findings suggest that partial spin reorientation can emerge without magnetic rare-earth ions and establish collective Raman phonon anomalies as a sensitive indirect probe of magnetic anisotropy changes in orthoferrites.
\end{abstract}


\vspace{0.8cm}

\begin{multicols}{2}
\section{Introduction}

Perovskite transition-metal oxides $AB$O$_3$ provide a versatile platform for investigating the coupling among charge, spin, orbital, and lattice degrees of freedom~\cite{Imada1998,TokuraNagaosa2000,rahman2023}. Their magnetic interactions are largely governed by superexchange through the $B$--O--$B$ bonds~\cite{Goodenough1955,Kanamori1959}, which are sensitive to the bond angles and cooperative tilting of the $B$O$_6$ octahedra~\cite{Glazer1972}. Consequently, chemical substitution at either cation site can modify the magnetic ground state by altering both the electronic configuration and the local structural environment.

Rare-earth orthoferrites $R$FeO$_3$ exemplify this interplay between crystal structure and magnetism~\cite{Treves1965,White1969}. They exhibit G-type antiferromagnetic ordering of Fe$^{3+}$ moments, with N\'eel temperatures of approximately 620--740~K~\cite{White1969,Eibschutz1967}. The Dzyaloshinskii-Moriya interaction produces a small canting of the antiferromagnetic sublattices, giving rise to weak ferromagnetism~\cite{Dzyaloshinsky1958,Moriya1960}. Although the exchange interaction stabilizes the antiferromagnetic order, the orientation of the N\'eel vector $\mathbf{G}$ is governed by comparatively weak magnetic anisotropy. In several orthoferrites, the temperature dependence of this anisotropy drives a spin-reorientation transition (SRT), commonly involving rotation of $\mathbf{G}$ from the $a$ axis ($\Gamma_4$) toward the $c$ axis ($\Gamma_2$)~\cite{HornerVarma1968,Yamaguchi1974,Belov1976}. Such transitions are frequently associated with interactions between the Fe and magnetic rare-earth sublattices~\cite{White1969,Belov1976}. The possibility of controlling the antiferromagnetic order parameter has stimulated interest in orthoferrites for ultrafast spin manipulation~\cite{Kimel2004,Schlauderer2019} and antiferromagnetic spintronics~\cite{Jungwirth2016,Baltz2018}. Chemical substitution provides an additional route to controlling the reorientation temperature and magnetic anisotropy. For example, spin reorientation near room temperature has been reported in Sm$_{0.5}$Y$_{0.5}$FeO$_3$~\cite{Mali2022}, while transition-metal substitution can modify the reorientation behavior, including re-entrant transitions in TbFe$_{1-x}$Mn$_x$O$_3$ and TbFe$_{0.5}$Cr$_{0.5}$O$_3$~\cite{Mandal2011,Fang2016,Mali2020}.

Among the orthoferrites, LaFeO$_3$ is particularly interesting because La$^{3+}$ is nonmagnetic and its N\'eel temperature is approximately 740~K~\cite{Koehler1957,Eibschutz1967}. Its magnetic structure is generally described by the $\Gamma_4$ configuration, with the N\'eel vector along the $a$ axis and the weak ferromagnetic moment along $c$~\cite{Scholl2000,Massa2023}. Unlike orthoferrites containing magnetic rare-earth ions, LaFeO$_3$ lacks the rare-earth--Fe magnetic coupling commonly responsible for spin reorientation, and no spontaneous reorientation has been established in the bulk compound~\cite{BelovKadomtseva1971,Belov1976}. Nevertheless, its magnetic anisotropy remains sensitive to structural perturbations, as demonstrated by strain-induced changes in the spin orientation of epitaxial films~\cite{Scholl2000}. This raises the question of whether modification of the Fe sublattice itself can induce a comparable response. Substitution with Ni offers a possible route because it introduces Fe--O--Ni exchange pathways and modifies the local electronic and structural environment. The LaFe$_{1-x}$Ni$_x$O$_3$ system exhibits composition-dependent structural, electronic, and magnetic properties~\cite{Falcon1997,Rao1975,Adachi2021,MAITY2026189520,maitylowNi,maitynih}. Previous studies have reported changes in coercivity, remanence, and magnetic ordering upon Ni substitution~\cite{Nguyen2019,Choudhary2005}. However, its influence on the orientation and temperature-dependent stability of the N\'eel vector remains insufficiently understood.

Raman spectroscopy provides a sensitive probe of spin-lattice coupling through deviations of phonon frequencies from their conventional anharmonic temperature dependence. Such deviations arise because lattice vibrations modulate magnetic exchange interactions and their associated spin correlations~\cite{Baltensperger1968,Granado1999,Laverdiere2006}. In orthoferrites, the Raman-active modes predominantly involve oxygen and rare-earth displacements, making them sensitive to structural distortions that influence the Fe--O--Fe exchange geometry~\cite{Weber2016}. However, the phonon response to spin reorientation is not universal. Anomalous temperature-dependent phonon behavior has been observed in NdFeO$_3$ and Sm$_{0.5}$Y$_{0.5}$FeO$_3$~\cite{Vilarinho2022,Mali2022}, whereas in SmFeO$_3$ the reorientation is more evident in the elastic response than in the optical phonon frequencies~\cite{Weber2022}. Related magnetic excitations also provide signatures of reorientation~\cite{Dubrovin2025}. Identifying such transitions from Raman measurements is therefore challenging, particularly when phonon anomalies are weak, mode dependent, or comparable to experimental uncertainties. Moreover, an anomalous frequency shift alone cannot uniquely distinguish a complete spin rotation from a partial, re-entrant rotation or fluctuations about the equilibrium spin orientation.

Here, we investigate the magnetic and lattice responses of LaFe$_{1-x}$Ni$_x$O$_3$ ($x=0$ and 0.1) using temperature-dependent magnetization and Raman spectroscopy. We analyze the correlated anomalous shifts of 12 Raman modes through principal component analysis and a global fitting approach, and develop a phenomenological spin-phonon coupling model to distinguish complete, partial, and incipient spin-reorientation scenarios. The combined results favor a partial, re-entrant reorientation near room temperature and reveal that Ni substitution modifies the characteristic temperature range and magnetic response. These findings suggest a route to tuning magnetic anisotropy in orthoferrites without relying on magnetic rare-earth ions.

\section{Experimetal}
Polycrystalline LaFe$_{0.9}$Ni$_{0.1}$O$_3$ was synthesized via a conventional sol--gel method using stoichiometric amounts of LaCl$_3\cdot$7H$_2$O (99\%, SRL), Fe(NO$_3$)$_3\cdot$9H$_2$O (98\%, LOBA Chemie), and NiCl$_2\cdot$6H$_2$O (97\%, LOBA Chemie). The precursors were dissolved in a water-ethanol mixture (1:2 by volume), maintaining a total solution volume of 30~mL and a metal-ion concentration of 0.167~M. Citric acid (0.010~mol, 99\%, SRL) was added as a chelating agent, followed by 1.5~mL of 5~wt.\% polyvinyl alcohol solution and 1~mL of ethylene glycol. The solution pH was adjusted to approximately 9 using aqueous ammonia and stirred continuously for 3~h. The resulting sol was heated at 150~$^\circ$C for 12~h to induce gel formation, followed by thermal treatment at 260~$^\circ$C for 6~h. The obtained powder was thoroughly ground and annealed at 1000~$^\circ$C for 12~h. For comparison, previously reported data for pristine LaFeO$_3$ were used as the undoped reference. Hereafter, LaFeO$_3$ and LaFe$_{0.9}$Ni$_{0.1}$O$_3$ are referred to as the pure and doped samples, respectively.

The crystalline structure and phase composition were examined by x-ray diffraction (XRD, Rigaku) using Co K$\alpha$ radiation ($\lambda\approx1.79$~\AA). Surface morphology and elemental distribution were investigated using field-emission scanning electron microscopy (FESEM, JEOL JSM-6500) coupled with energy-dispersive x-ray spectroscopy (EDS). Microstructural features and crystallinity were further examined using high-resolution transmission electron microscopy (HRTEM, Thermo Scientific) operated at 300~kV. Temperature-dependent Raman spectra were recorded over 200--400~K using a Renishaw inVia Reflex confocal Raman spectrometer with an excitation wavelength of 830~nm. Magnetization measurements were performed using a vibrating sample magnetometer (VSM, Cryogenic Ltd., UK) under zero-field-cooled (ZFC) and field-cooled (FC) protocols.

\section{Results and Discussions}

\subsection{Structural analysis}
\label{sec:structure}

\begin{figure}[H]
\centering
\includegraphics[width=\columnwidth]{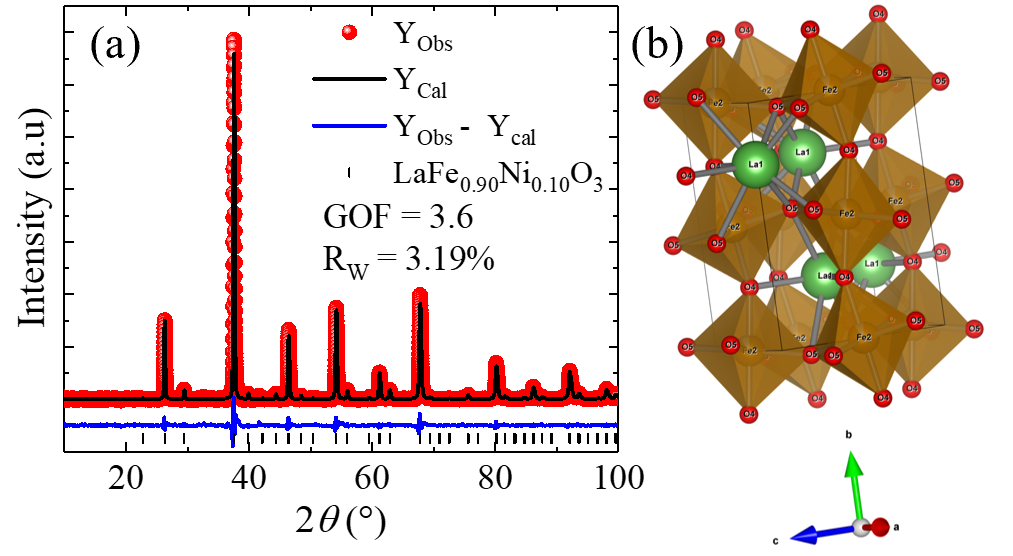}
\caption{(a) Rietveld refinement of the PXRD pattern of LaFe$_{0.9}$Ni$_{0.1}$O$_3$: observed and calculated profiles, Bragg positions and difference curve. (b) Refined crystal structure ($Pnma$).}
\label{fig:xrd}
\end{figure}

The phase purity and crystal structure of LaFe$_{0.9}$Ni$_{0.1}$O$_3$ were determined by powder x-ray diffraction (PXRD) and Rietveld refinement with GSAS-II~\cite{Toby2013} [Fig.~\ref{fig:xrd}(a)]. All reflections are described by the orthorhombic perovskite structure, space group $Pnma$ (No.~62), and the difference curve shows the systematic deviations. The refined lattice parameters and selected bond lengths and angles are listed in Table~\ref{tab:structure}, together with those of our pristine LaFeO$_3$~\cite{MAITY2026189520} and of LaFeO$_3$ from neutron diffraction~\cite{dixon2015}.

The lattice constants of LaFe$_{0.9}$Ni$_{0.1}$O$_3$, $a=5.55693$, $b=7.85108$ and $c=5.54864$~\AA\ ($V=242.08$~\AA$^3$), are close to those of bulk LaFeO$_3$ ($V=241.94$~\AA$^3$ at 50~K~\cite{dixon2015}). The cell of our pristine LaFeO$_3$~\cite{MAITY2026189520}, refined separately, is about 2\% smaller ($V=237.18$~\AA$^3$). Because low-spin Ni$^{3+}$ is smaller than high-spin Fe$^{3+}$~\cite{Falcon1997}, Ni substitution is not expected to expand the lattice, and we do not attribute this difference to Ni. The B--O--B angles of the substituted sample, $160.00^\circ$ (B--O1--B) and $155.44^\circ$ (B--O2--B), are 1--3$^\circ$ larger than in pristine LaFeO$_3$, and its two B--O2 distances, 1.996 and 2.022~\AA, are much closer to each other than those reported there (1.871 and 2.146~\AA), which suggests a less distorted BO$_6$ network. X-ray diffraction determines the oxygen positions less precisely than those of the cations, so this comparison is indicative only. Angles of 155--160$^\circ$ set the Fe--O--Fe superexchange geometry, to which the Raman modes of Sec.~\ref{sec:raman} couple.

\begin{center}
\captionof{table}{Structural parameters of LaFe$_{0.9}$Ni$_{0.1}$O$_3$ ($x=0.1$, this work), pristine LaFeO$_3$~\cite{MAITY2026189520}, and LaFeO$_3$ at 50~K from neutron diffraction~\cite{dixon2015}. All structures are described by the $Pnma$ space group. B denotes the Fe or mixed Fe/Ni site. The O2 site gives two inequivalent B--O2 bond lengths.}
\label{tab:structure}
\resizebox{\columnwidth}{!}{%
\begin{tabular}{lccc}
\hline\hline
Parameters & $x=0$ & $x=0.1$ & $x=0$ (50~K) \\
\hline
$a$ (\AA) & 5.52197 & 5.55693 & 5.56275 \\
$b$ (\AA) & 7.78523 & 7.85108 & 7.8416 \\
$c$ (\AA) & 5.51714 & 5.54864 & 5.54642 \\
$V$ (\AA$^3$) & 237.18 & 242.08 & 241.94 \\
B--O1--B ($^\circ$) & 158.71 & 160.00 & ~155 \\
B--O2--B ($^\circ$) & 152.58 & 155.44 & ~159.5 \\
B--O1 (\AA) & 1.980 & 1.993 & ~2.025 \\
B--O2 (\AA) & 1.871, 2.146 & 1.996, 2.022 & ~2.005,~2.01 \\
\hline\hline
\end{tabular}%
}
\end{center}

\subsection{Microstructure}
\label{sec:microstructure}

\begin{figure}[H]
\centering
\includegraphics[width=\columnwidth]{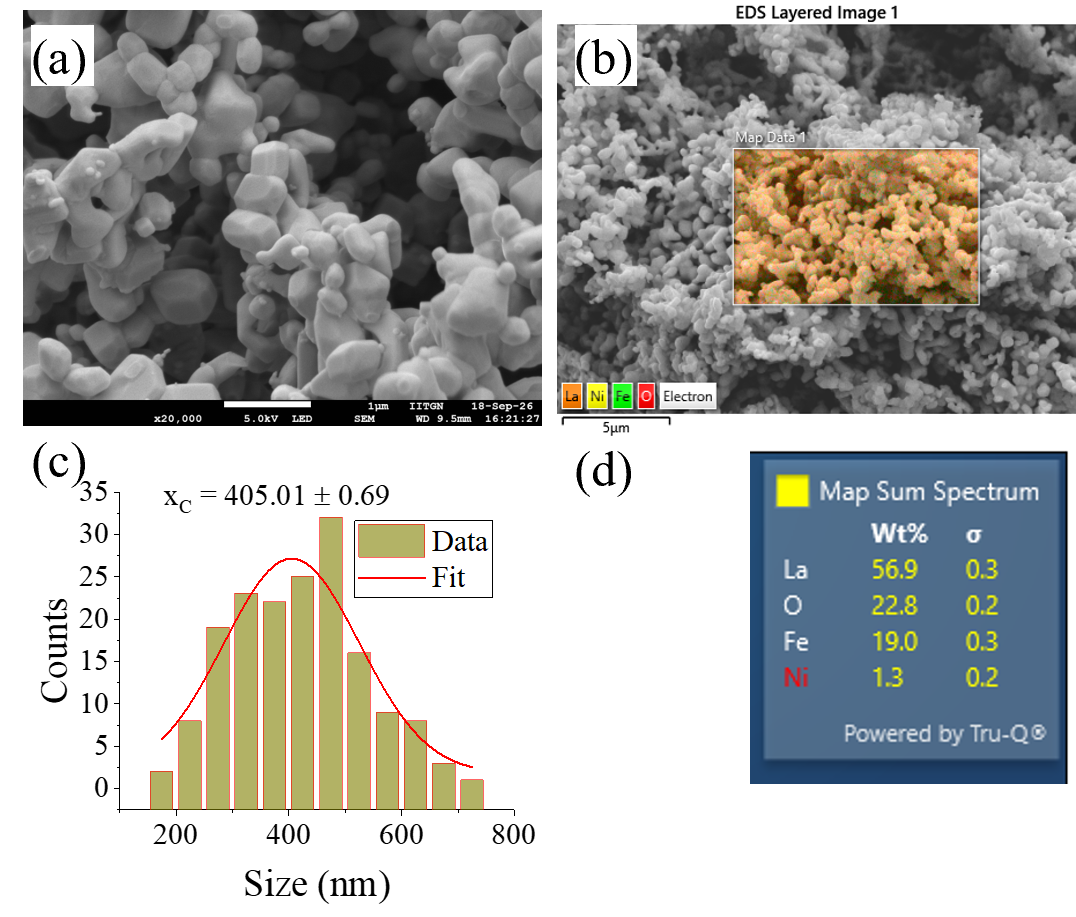}
\caption{LaFe$_{0.9}$Ni$_{0.1}$O$_3$: (a) SEM image, (b) EDS elemental maps,
(c) particle-size distribution with a normal fit, and (d) elemental weight fractions.}
\label{fig:sem}
\end{figure}

The SEM image of LaFe$_{0.9}$Ni$_{0.1}$O$_3$ [Fig.~\ref{fig:sem}(a)] shows well-faceted polyhedral particles with sharp edges and distinct grain boundaries, together with some agglomeration from sintering during annealing. The particle size, taken as the side of a square of equal projected area and measured with ImageJ for \val{~170} particles, follows a normal distribution with a mean of about 405~nm and a standard deviation of \val{0.69}~nm [Fig.~\ref{fig:sem}(c)]. This is larger than the 337~nm of pristine LaFeO$_3$ prepared under the same conditions~\cite{MAITY2026189520}, which suggests that Ni promotes grain growth during annealing. The EDS maps [Fig.~\ref{fig:sem}(b)] show La, Fe, Ni and O throughout the imaged region. The measured weight fractions [Fig.~\ref{fig:sem}(d)] correspond to a cation ratio of LaFe$_{0.94}$Ni$_{0.06}$O$_3$, somewhat below the nominal Ni content. Standardless EDS is affected by the interaction volume, matrix effects, surface roughness and its low sensitivity to oxygen, so this composition is semiquantitative, and the nominal one is used throughout~\cite{Newbury2013}.

\begin{figure}[H]
\centering
\includegraphics[width=\columnwidth]{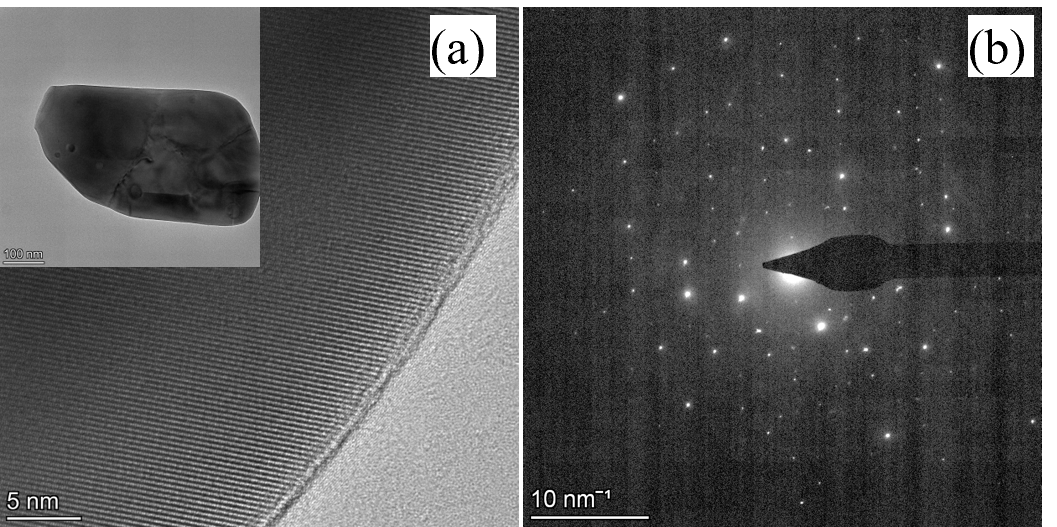}
\caption{LaFe$_{0.9}$Ni$_{0.1}$O$_3$: (a) high-resolution TEM image showing lattice
fringes and (b) SAED pattern.}
\label{fig:tem}
\end{figure}

For transmission electron microscopy, the powder was dispersed in isopropanol by ultrasonication and drop-cast onto a carbon-coated grid. The high-resolution image [Fig.~\ref{fig:tem}(a)] shows continuous lattice fringes across large parts of the particles, with a spacing of \val{0.36}~nm, close to that of the \val{(111)} planes of the $Pnma$ structure (0.351~nm). The selected-area electron diffraction (SAED) pattern [Fig.~\ref{fig:tem}(b)] consists of sharp spots without rings, from a single crystallite. Together with PXRD and SEM, these images show well-crystallized particles with long-range lattice order.

\subsection{Magnetization}
\label{sec:magnetic}

\begin{figure}[H]
\centering
\includegraphics[width=\columnwidth]{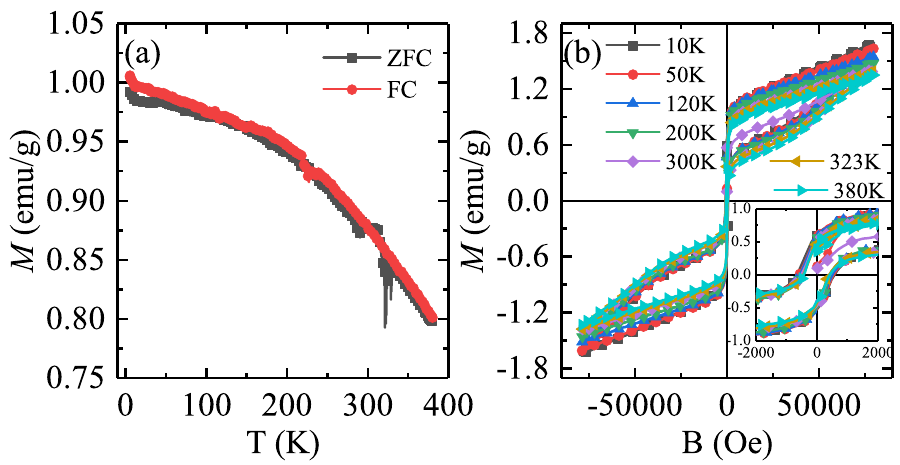}
\caption{LaFe$_{0.9}$Ni$_{0.1}$O$_3$: (a) ZFC and FC magnetization at 5~kOe and (b) magnetization loops at different temperatures.}
\label{fig:mhmt}
\end{figure}

Figure~\ref{fig:mhmt}(a) shows the magnetization of LaFe$_{0.9}$Ni$_{0.1}$O$_3$ measured between 3 and 380~K at 5~kOe after zero-field cooling (ZFC) and on field cooling (FC). $M(T)$ changes only weakly over this range, and the ZFC and FC curves separate slightly without a cusp, indicating weak magnetic irreversibility rather than an ordering or freezing transition. The magnetization loops [Fig.~\ref{fig:mhmt}(b)] are hysteretic at all temperatures, with coercive fields of 435--650~Oe and remanence of 0.45--0.60~emu\,g$^{-1}$, on a nearly linear, nonsaturating high-field background. This is the response of a weak ferromagnetic moment in an antiferromagnetic matrix, as expected for the canted G-type order of an orthoferrite, in which the Dzyaloshinskii--Moriya interaction cants the two sublattices~\cite{Dzyaloshinsky1958,Moriya1960,MAITY2026189520,Autieri2025StaggeredDMI}. The loops are also constricted near zero field (wasp-waisted), as in the LaFeO$_3$ and LaFe$_{1-x}$Ni$_x$O$_3$ samples of our earlier study~\cite{MAITY2026189520}. Such a constriction arises when the anisotropy-locked canted moment, which reverses only at higher fields, coexists with a softer contribution that reverses near zero field. The softer contribution can come from Ni-related exchange~\cite{MAITY2026189520}, from uncompensated surface spins and from the distribution of particle sizes and anisotropies~\cite{Ahmadvand2010ExchangeBias,Nguyen2012SizeEffect,KOFERSTEIN20131}; a minority magnetic phase below the PXRD detection limit would contribute in the same way.

\begin{figure*}[t]
\centering
\includegraphics[width=0.9\textwidth]{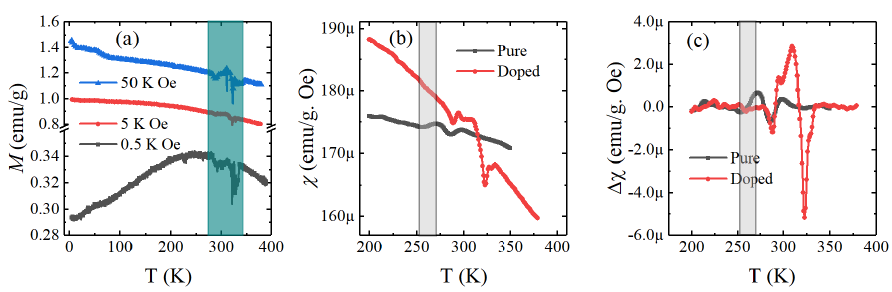}
\caption{(a) ZFC magnetization of LaFe$_{0.9}$Ni$_{0.1}$O$_3$ at 0.5, 5 and 50~kOe. (b) ZFC susceptibility of LaFeO$_3$ (black), offset by \val{$0.78\times10^{-4}$}~emu\,g$^{-1}$\,Oe$^{-1}$ for clarity, and of LaFe$_{0.9}$Ni$_{0.1}$O$_3$ (red). (c) Relative deviation $\delta\chi$ from a cubic background fitted outside the anomaly. Dashed lines in (b) and (c) mark 255--268~K, the feature of trace $\alpha$-Fe$_2$O$_3$.}
\label{fig:dchi}
\end{figure*}

Within the ordered state the susceptibility shows a broad, nonmonotonic anomaly near room temperature. In LaFe$_{0.9}$Ni$_{0.1}$O$_3$ it appears at the same temperature in ZFC measurements at 0.5, 5 and 50~kOe [Fig.~\ref{fig:dchi}(a)], so it is intrinsic and its temperature does not depend on field. Figure~\ref{fig:dchi}(b) compares the ZFC susceptibility, $\chi=M/H$, of the two samples. The anomaly extends over about 250--310~K in LaFeO$_3$ and 280--330~K in LaFe$_{0.9}$Ni$_{0.1}$O$_3$. To isolate it, a cubic polynomial $\chi_{\mathrm{bg}}(T)$ was fitted outside 245--315~K ($x=0$) and 275--340~K ($x=0.1$), and Fig.~\ref{fig:dchi}(c) shows the relative deviation $\delta\chi=(\chi-\chi_{\mathrm{bg}})/\chi_{\mathrm{bg}}$. These intervals, fixed from $\chi(T)$ alone, also serve as the magnetic windows of the Raman analysis (Sec.~\ref{sec:raman}).

In LaFeO$_3$, $\delta\chi$ reaches $-0.73\%$ at 286~K and recovers to $+0.43\%$ near 299~K. A weaker positive feature near 264~K lies at the Morin transition of $\alpha$-Fe$_2$O$_3$~\cite{Morin1950}, a trace of which is detected by Raman scattering (Sec.~\ref{sec:raman}). In LaFe$_{0.9}$Ni$_{0.1}$O$_3$ the anomaly is shifted to higher temperature and is stronger: $\delta\chi$ rises to $+1.7\%$ near 312~K and falls to $-3.1\%$ at 322~K, a negative excursion about four times larger than in LaFeO$_3$.

These features lie far below $T_{\mathrm N}$ and are therefore not related to the onset of long-range order. A spin reorientation~\cite{White1969,Belov1976,Mandal2011}, a change of the canting~\cite{Treves1965,Zhao2016}, domain-wall rearrangement~\cite{Bazaliy2010}, blocking of magnetic clusters~\cite{Bean1959}, uncompensated surface spins~\cite{Kodama1997,Ahmadvand2010ExchangeBias} and magnetoelastic effects~\cite{Vilarinho2022,Weber2022} could all produce such features. In LaFe$_{0.9}$Ni$_{0.1}$O$_3$ the anomaly persists at 50~kOe, about a hundred times the coercive field, and its temperature does not change between 0.5 and 50~kOe; together with the absence of a cusp in the ZFC magnetization, this argues against domain-wall effects and the blocking of clusters, but the dc susceptibility alone cannot distinguish among the remaining mechanisms. Unlike the 264~K feature, the main anomaly occurs in both samples, and its shift and enhancement with Ni suggest that it is governed by the exchange paths and the magnetic anisotropy of the Fe sublattice, which Ni on the Fe site modifies. We therefore examine it with temperature-dependent Raman spectroscopy. We therefore examine it with temperature-dependent Raman spectroscopy, which probes the whole volume of the grains (Sec.~\ref{sec:raman}) and whose phonon frequencies depend on the spin correlations through spin--phonon coupling~\cite{Granado1999}. A change of the spin structure throughout the grains can therefore shift them~\cite{Vilarinho2022,Mali2022}, whereas a change confined to the surface, a layer of about 1~nm that makes up less than 2\% of a 340--405~nm grain, contributes to the measured frequencies only in proportion to that volume.

\begin{figure*}[!t]
\centering
\includegraphics[width=0.9\textwidth]{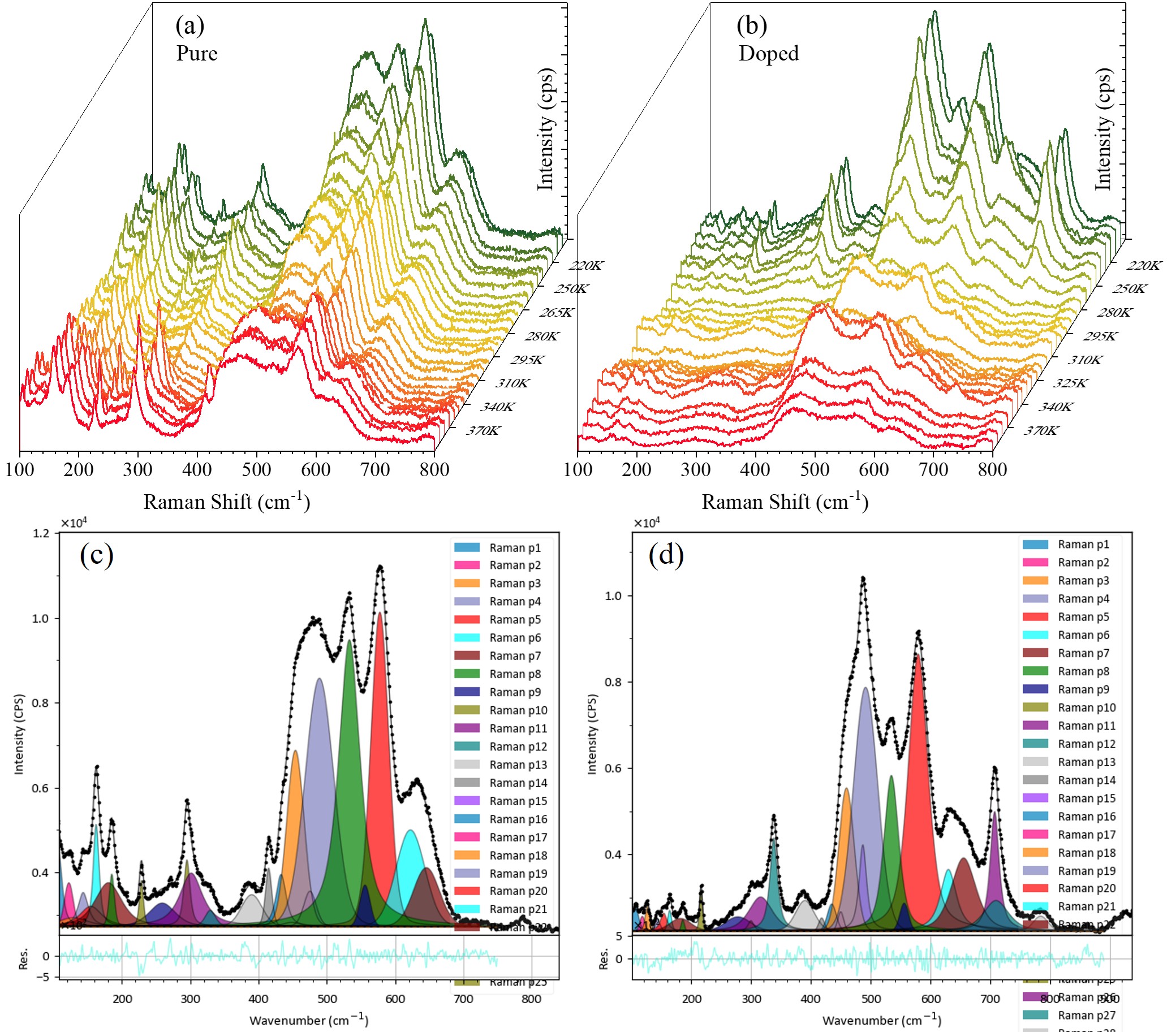}
\caption{Raman spectra of (a) LaFeO$_3$ and (b) LaFe$_{0.9}$Ni$_{0.1}$O$_3$ between 200 and 400~K (green to red with increasing temperature). (c), (d) Decomposition of the spectra at 200~K.}
\label{fig:raman1}
\end{figure*}

\subsection{Raman spectroscopy}
\label{sec:raman}

Raman spectra of both samples were recorded between 200 and 400~K on warming, after cooling to 200~K, with 830~nm (1.49~eV) excitation and a laser power of \val{0.8}~mW at the sample. The temperature step was 10~K, reduced to 5~K around the magnetic anomaly, and each spectrum was accumulated three times for 60~s. The excitation lies below the optical gap of LaFeO$_3$ (about 2.3~eV~\cite{Scafetta2014}), so the spectra are nonresonant and probe the whole volume of the grains. The peaks broaden and their heights decrease gradually with temperature [Figs.~\ref{fig:raman1}(a),(b)]. Each spectrum was decomposed with KherveFitting into pseudo-Voigt components (SGL) on a constant background, with the same model at all temperatures. Figures~\ref{fig:raman1}(c),(d) show the fits at 200~K, and Table~\ref{tab:raman} lists the component positions.

Orthorhombic LaFeO$_3$ ($Pnma$, $Z=4$) has 24 Raman-active modes, $\Gamma_{\mathrm{Raman}}=7A_g+5B_{1g}+7B_{2g}+5B_{3g}$. Fe occupies the centrosymmetric $4b$ site and does not move in these modes, which involve only La and O displacements~\cite{Weber2016}. They therefore couple to the Fe spins through the Fe--O--Fe bond angles and Fe--O bond lengths (Appendix~\ref{app:phonon}). Weber \textit{et al.} observed 19 of the 24 modes in polycrystalline LaFeO$_3$ at 80~K and assigned them with DFT~\cite{Weber2016}. Comparing our 25 components with these modes, we sort them into three classes (Table~\ref{tab:raman}). Class~A contains components that agree with modes observed in Ref.~\cite{Weber2016}, allowing for the shift between 80 and 200~K. Class~B contains tentative assignments, to modes predicted but not observed there or observed with a larger offset. Class~C contains components without a clear first-order counterpart.

The 12 Class-A components account for 14 of the observed modes, because p16 contains three modes between 428 and 436~cm$^{-1}$ that cannot be separated without polarized measurements. With the six Class-B components, at most 20 of the 24 allowed modes are represented, consistent with the mode count, so the Class-C components are not first-order phonons of LaFeO$_3$. p2 and p7 describe broad shoulders, and p13, p20 and p23 lie near sums of first-order frequencies, such as $2A_g(3)$, $A_g(3)+B_{1g}(3)$ and $A_g(4)+A_g(5)$, and are tentatively attributed to second-order or disorder-activated scattering. In LaFeO$_3$, the sharp p25 component (228.6~cm$^{-1}$) and part of p19 (489~cm$^{-1}$) lie close to the $A_{1g}$ modes of $\alpha$-Fe$_2$O$_3$ at 226 and 498~cm$^{-1}$~\cite{deFaria1997}. Since hematite scatters much more strongly than LaFeO$_3$, these weak features indicate a trace of $\alpha$-Fe$_2$O$_3$ below the PXRD detection limit.

\begin{table*}[t]
\caption{Raman components of LaFeO$_3$ ($x=0$) and LaFe$_{0.9}$Ni$_{0.1}$O$_3$ ($x=0.1$) at 200~K, with the frequencies measured at 80~K in Ref.~\cite{Weber2016} (calculated values in parentheses) and the assignments. Class~A (bold): clear counterpart among the modes observed in Ref.~\cite{Weber2016}; these 12 components are used in the analysis. Class~B: tentative assignment. Class~C: weak, overlapping, second-order or secondary-phase component.}
\label{tab:raman}
\begin{tabular}{cccclc}
\hline\hline
Peak & $x=0$ (cm$^{-1}$) & $x=0.1$ (cm$^{-1}$) & Ref.~\cite{Weber2016} (cm$^{-1}$) & Assignment, main motion & Class \\
\hline
p1  & 105.5 & 106.3 & 105.5 (103) & $B_{2g}(1)$, La $R(z)$ & \textbf{A} \\
p2  & 122.4 & 118.3 & --          & low-frequency component & C \\
p3  & 128.6 & 127.0 & 135.3 (127) & $A_g(2)$, La $R(z)$, out-of-phase & B \\
p4  & 143.4 & 143.0 & 143.0 (144) & $B_{2g}(2)$, La $R(x)$ & \textbf{A} \\
p5  & 154.5 & 154.2 & -- (148)    & $B_{1g}(2)$, [010] FeO$_6$ rotation & B \\
p6  & 162.3 & 164.0 & 166.5 (172) & $B_{2g}(3)$, [101] FeO$_6$ rotation & \textbf{A} \\
p7  & 180.2 & 182.0 & --          & broad component under $A_g(3)$ & C \\
p8  & 185.3 & 186.0 & 186.6 (183) & $A_g(3)$, [010] FeO$_6$ rotation, in-phase & \textbf{A} \\
p25 & 228.6 & 216.4 & --          & $\alpha$-Fe$_2$O$_3$ $A_{1g}$ ($x=0$) & C \\
p9  & 259.0 & 278.0 & 274.2 (273) & $A_g(4)$, O(1) motion in the $x$--$z$ plane & B \\
p10 & 295.3 & 299.8 & 302.8 (306) & $A_g(5)$, [101] FeO$_6$ rotation & B \\
p11 & 301.4 & 316.0 & 316.8 (300) & $B_{3g}(2)$, O(1)--Fe--O(2) & B \\
p12 & 329.4 & 337.9 & 338.1 (328) & $B_{1g}(3)$, [010] FeO$_6$ rotation, out-of-phase & \textbf{A} \\
p13 & 390.1 & 390.0 & --          & no first-order counterpart & C \\
p14 & 414.8 & 418.0 & 416.8 (401) & $B_{2g}(5)$, O--Fe--O bending & \textbf{A} \\
p16 & 433.1 & 434.9 & 428.6--436.0 & $A_g(7)$, $B_{3g}(4)$, $B_{3g}(3)$ (unresolved) & \textbf{A} \\
p17 & 448.1 & 449.8 & 442.3 (425) & $B_{1g}(4)$, Fe--O(2) stretching & \textbf{A} \\
p18 & 454.1 & 459.5 & 449.8 (433) & $A_g(6)$, Fe--O(2) stretching & \textbf{A} \\
p15 & 475.4 & 486.8 & -- (481)    & $B_{2g}(6)$, O--Fe--O bending & B \\
p19 & 489.2 & 491.4 & --          & $\alpha$-Fe$_2$O$_3$ $A_{1g}$ ($\sim$498) & C \\
p23 & 533.1 & 534.6 & --          & no first-order counterpart & C \\
p24 & 555.9 & 555.9 & 560.9 (584) & $B_{1g}(5)$, Fe--O(1) stretching & \textbf{A} \\
p20 & 577.6 & 579.2 & --          & no first-order counterpart & C \\
p21 & 622.4 & 629.8 & 625.1 (622) & $B_{2g}(7)$, Fe--O(2) stretching & \textbf{A} \\
p22 & 645.8 & 655.2 & 641.9 (650) & $B_{3g}(5)$, FeO$_6$ breathing & \textbf{A} \\
\hline\hline
\end{tabular}
\end{table*}

All quantitative analyses below use only the 12 Class-A modes. They were selected for the reliability of their assignment before their temperature dependence was examined, so the selection does not favor modes with large anomalies.

\begin{figure*}[!t]
\centering
\includegraphics[width=0.95\textwidth]{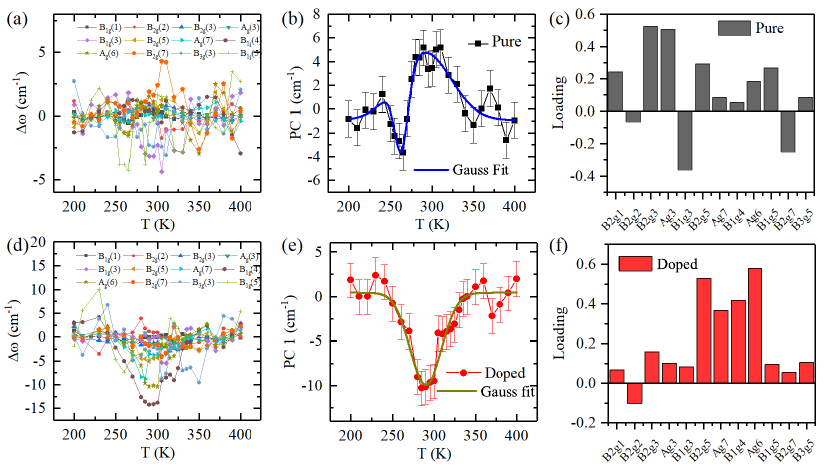}
\caption{(a), (d) Anomalous shifts of the 12 Class-A modes of LaFeO$_3$ and LaFe$_{0.9}$Ni$_{0.1}$O$_3$, relative to the anharmonic baseline, Eq.~(\ref{eq:anharmonic}). (b), (e) First principal component PC1 of the noise-scaled shifts (in units of the noise), with Gaussian fits (two Gaussians for LaFeO$_3$). (c), (f) PC1 loadings.}\label{fig:pca}
\end{figure*}

Without coupling to other excitations, the frequency of an optical phonon follows its cubic anharmonic decay into two phonons~\cite{balkanski28}. With Raman shifts in cm$^{-1}$, the baseline of mode $i$ is
\begin{equation}
\omega_i^{\mathrm{anh}}(T)=\omega_{0,i}+C_i\left[1+\frac{2}{e^{x_i}-1}\right],
\ x_i=\frac{hc\,\omega_{0,i}}{2k_{\mathrm B}T},
\label{eq:anharmonic}
\end{equation}
where $\omega_{0,i}$ is the zero-temperature frequency and $C_i$ the anharmonic coefficient. Both were fitted only to the temperatures outside the magnetic windows, 250--315~K ($x=0$) and 275--340~K ($x=0.1$), which were fixed from $\chi(T)$ alone (Sec.~\ref{sec:magnetic}), so that the anomaly does not influence its own baseline. Figures~\ref{fig:pca}(a) and \ref{fig:pca}(d) show the anomalous shifts, $\Delta\omega_i=\omega_i-\omega_i^{\mathrm{anh}}$. In both samples the frequencies leave the baseline only within a finite interval around the magnetic anomaly and return to it on both sides.

To extract the response common to all modes without assuming its shape, we apply principal component analysis~\cite{Jolliffe2016} to the noise-scaled shifts $\Delta\omega_i/\sigma_i$, where $\sigma_i$ is the rms of $\Delta\omega_i$ outside the window (Appendix~\ref{appBalk:reduction}). The first principal component, PC1 [Figs.~\ref{fig:pca}(b),(e)], is the largest correlated variation of the 12 modes, and its loadings [Figs.~\ref{fig:pca}(c),(f)] give the contribution of each mode. Independently, the 12 modes were fitted simultaneously with a common profile, $\Delta\omega_i(T)=y_{0,i}+A_i\,g(T)$, with an offset $y_{0,i}$ and an amplitude $A_i$ for each mode [Eq.~(\ref{eq:A_global})]. This global fit separates the shared temperature dependence from the size and sign of each mode's response.

In LaFe$_{0.9}$Ni$_{0.1}$O$_3$, PC1 carries 62\% of the variance and has a single broad maximum near 290~K. Its largest loadings come from the oxygen stretching and bending modes A$_g$(6), B$_{2g}$(5) and B$_{1g}$(4) and the unresolved A$_g$(7)/B$_{3g}$ band, all of which soften within the window. In LaFeO$_3$, PC1 carries 44\% of the variance and has two parts. A broad component near 290~K is dominated by the octahedral-rotation modes B$_{2g}$(3) and A$_g$(3), which harden. A narrow component near 262~K coincides with the $\chi$ feature at 264~K and with the Morin transition of $\alpha$-Fe$_2$O$_3$~\cite{Morin1950}. It appears weakly ($-0.5$ to $-0.9$~cm$^{-1}$) also in La-dominated and rotation modes that overlap no hematite line, which suggests that the Morin transition of the trace $\alpha$-Fe$_2$O$_3$ slightly perturbs the perovskite lattice, for instance through strain at the phase boundaries.

Gaussian fits of PC1 (two Gaussians for $x=0$) and the global fit with a Gaussian $g(T)$ of common centre $T_c$ and width $\sigma$ agree within their errors (Table~\ref{tab:summary}). The broad response is centred at the same temperature, about 290~K, in both samples, but is narrower in LaFe$_{0.9}$Ni$_{0.1}$O$_3$ ($\sigma\approx19$~K) than in LaFeO$_3$ ($\sigma\approx32$~K); LaFeO$_3$ shows in addition a narrow component at 262~K ($\sigma\approx8$~K). Its centre agrees with the steepest rise of $\chi(T)$ at 264~K, measured without illumination, which limits laser heating in LaFeO$_3$ to a few kelvin. The Gaussian serves only to locate the response; its shape is analyzed in Sec.~\ref{sec:comparison}.

The response is not a uniform change of volume: the shifts have opposite signs in different modes, and the largest, about 2\% for A$_g$(6) in LaFe$_{0.9}$Ni$_{0.1}$O$_3$, would require a volume change of order 1\%. It is strongest in modes that change the Fe--O--Fe angle or the Fe--O length, which control the superexchange and the Dzyaloshinskii--Moriya interaction. Because $T_c$ does not depend on the analysis, it can be compared directly with the magnetization: in LaFeO$_3$ the broad component coincides, within its uncertainty, with the $\delta\chi$ minimum at 286~K, whereas in LaFe$_{0.9}$Ni$_{0.1}$O$_3$ it lies about 33~K below the $\delta\chi$ minimum at 323~K. Sections~\ref{sec:theory} and \ref{sec:comparison} show that both positions follow from a partial rotation of the N\'eel vector.

\subsection{Theoretical analysis}
\label{sec:theory}

The results establish a magnetic crossover within the ordered state of both samples, to which the lattice responds through the Fe--O--Fe exchange geometry. In this section we set out a minimal model of the crossover and derive the phonon response expected for each possible pathway (Fig.~\ref{fig:theory}). In Sec.~\ref{sec:comparison} these predictions are compared with the Raman index and the first principal component of both samples (Fig.~\ref{fig:fits}). The analysis of the phonon data is described in Appendix~\ref{app:reduction}, and the derivations in Appendices~\ref{app:landau} and \ref{app:phonon}.

\subsubsection{Spin orientation and its anisotropy}

Four observations constrain the crossover. It occurs about 450~K below $T_{\mathrm N}$, so it is not an ordering transition; the samples are single-phase and well crystallized; its temperature does not depend on field and shows no ZFC--FC bifurcation, so it is not glassy; and the phonon anomaly is confined to the same interval and is strongest in modes that modify the Fe--O--Fe geometry. Within the ordered state of an orthoferrite, the quantity that can change in this way is the orientation of the N\'eel vector $\mathbf G$ relative to the crystal axes. With $\theta$ the angle between $\mathbf G$ and the $a$ axis ($\theta=0$: $\Gamma_4$; $\theta=\pi/2$: $\Gamma_2$), the orientation-dependent free energy is~\cite{HornerVarma1968,Belov1976}
\begin{equation}
\Phi(\theta,T)=K_1(T)\sin^2\theta+K_2\sin^4\theta ,
\label{eq:Phi}
\end{equation}
where $K_1$ is an effective anisotropy containing single-ion, anisotropic-exchange and Dzyaloshinskii-Moriya contributions [Eq.~(\ref{eq:A_K1K2})]. Minimizing $\Phi$ with respect to $s=\sin^2\theta$ gives [Eq.~(\ref{eq:A_s})]
\begin{equation}
s=-\frac{K_1}{2K_2},\qquad 0\le s\le1 ,
\label{eq:s_eq}
\end{equation}
so that $\mathbf G\parallel a$ ($\Gamma_4$) for $K_1\ge0$, $\mathbf G$ lies in the $ac$ plane ($\Gamma_{24}$) for $-2K_2<K_1<0$, and $\mathbf G\parallel c$ ($\Gamma_2$) for $K_1\le-2K_2$. For $K_2>0$ the spins rotate continuously; for $K_2<0$ they jump (Morin type). In orthoferrites that reorient in the $ac$ plane, $K_1$ changes sign almost linearly with temperature, whereas $K_2$ is positive and nearly constant, $K_2\approx3\times10^4$~erg~cm$^{-3}$ in TmFeO$_3$ and HoFeO$_3$~\cite{Belov1976}. We adopt the same description.

The angular stiffness $\kappa_\theta=\partial^2\Phi/\partial\theta^2$ [Eq.~(\ref{eq:A_kappa_general})] sets the thermal orientational fluctuations of a correlated region of volume $V$:
\begin{equation}
\kappa_\theta=
\begin{cases}
2K_1 & (\Gamma_4),\\
2K_2\sin^22\theta & (\Gamma_{24}),
\end{cases}
\
\left\langle\delta\theta^2\right\rangle=\frac{k_{\mathrm B}T}{V\kappa_\theta}.
\label{eq:stiffness}
\end{equation}
The spins become soft whenever $K_1\to0^+$, even without a static rotation, and at both edges of a rotation, where the quasi-ferromagnetic magnon softens, as observed in Sm$_{0.55}$Tb$_{0.45}$FeO$_3$~\cite{Dubrovin2025}.

\begin{figure*}[t]
\centering
\includegraphics[width=0.99\textwidth]{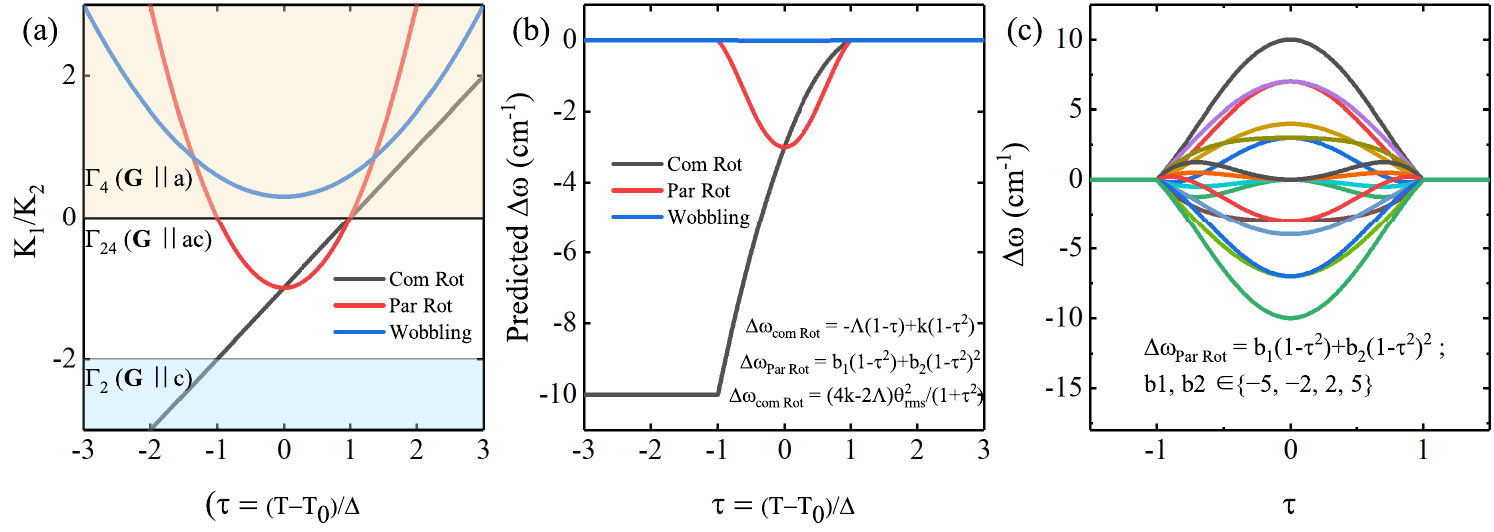}
\caption{Model predictions for the three spin-reorientation scenarios. (a) $K_1/K_2$ versus reduced temperature $\tau=(T-T_0)/\Delta$ for complete rotation (black), partial rotation (red), and incipient reorientation, i.e., angular wobbling (blue). (b) Corresponding phonon shifts $\Delta\omega$ from Eqs.~(\ref{eq:CR})--(\ref{eq:IR}), with $\Lambda=5$~cm$^{-1}$, $\kappa=2$~cm$^{-1}$, $s_{\max}=0.5$ ($b_1=-1$, $b_2=-2$~cm$^{-1}$), and $\theta_{\mathrm{rms}}=20^\circ$. (c) Partial-rotation shift [Eq.~(\ref{eq:PR})] for all 16 combinations $(b_1,b_2)\in\{\pm2,\pm5\}^2$~cm$^{-1}$.}\label{fig:theory}
\end{figure*}

Expanding $K_1(T)$ to second order about a characteristic temperature $T_0$ [Eq.~(\ref{eq:A_expansion})] gives three scenarios, each with one temperature scale $\Delta$. In the reduced temperature $\tau=(T-T_0)/\Delta$,
\begin{equation}
\frac{K_1}{K_2}=
\begin{cases}
\tau-1 & \text{(complete rotation)},\\
2s_{\max}\left(\tau^2-1\right) & \text{(partial rotation)},\\
k_{\min}\left(1+\tau^2\right) & \text{(incipient reorientation)},
\end{cases}
\label{eq:scenarios}
\end{equation}
which is plotted as a function of $\tau$ in Fig.~\ref{fig:theory}(a). In a complete rotation $K_1$ falls through the window, from zero at its upper edge ($\tau=+1$) to $-2K_2$ at its lower edge ($\tau=-1$); then $s=(1-\tau)/2$, and the spins turn from $a$ to $c$ on cooling [Eq.~(\ref{eq:A_complete})]. In a partial rotation $K_1$ dips below zero and recovers; then $s=s_{\max}(1-\tau^2)$, and the spins leave the $a$ axis at $\tau=+1$, turn to $\theta_{\max}=\arcsin\sqrt{s_{\max}}$ at $\tau=0$ and return at $\tau=-1$ [Eq.~(\ref{eq:A_partial})]. Such re-entrant reorientations occur in substituted orthoferrites; in TbFe$_{0.5}$Cr$_{0.5}$O$_3$, for example, the spins switch from $\Gamma_2$ to $\Gamma_4$ at 257~K and back at 190~K~\cite{Mali2020}. In an incipient reorientation $K_1$ has a positive minimum, $k_{\min}K_2$, at $T_0=T^*$. The spins never turn, $s=0$, but by Eq.~(\ref{eq:stiffness}) they fluctuate most at $T^*$, with $\langle\delta\theta^2\rangle\simeq\theta_{\mathrm{rms}}^2/(1+\tau^2)$ [Eq.~(\ref{eq:A_incipient})].

\subsubsection{Coupling of the spin orientation to the Raman phonons}

Fe occupies an inversion centre and does not move in Raman-active modes. The phonons displace the oxygen ions that mediate the Fe--O--Fe interactions and thereby modulate the anisotropy. To second order in the phonon coordinate and measured from $\Gamma_4$, the anomalous shift of mode $i$ is
[Eq.~(\ref{eq:A_shift_s})]
\begin{equation}
\Delta\omega_i(T)=-2\Lambda_i\langle s\rangle+4\kappa_i\langle s(1-s)\rangle ,
\label{eq:shift}
\end{equation}
where $\Lambda_i$ and $\kappa_i$ measure how the phonon modulates the second- and fourth-order anisotropy [Eq.~(\ref{eq:A_couplings})], and the brackets denote the average over the static orientation and its fluctuations. The isotropic spin-phonon coupling~\cite{Granado1999} varies smoothly far below $T_{\mathrm N}$ and is contained in the anharmonic baseline. The couplings differ from mode to mode and may have either sign, but the temperature dependence enters only through $\langle s\rangle$ and $\langle s(1-s)\rangle$, which are the same for all modes. Inserting $s(\tau)$ of each scenario gives explicit responses (Appendix~\ref{app:phonon}). For a complete rotation,
\begin{equation}
\Delta\omega_{\mathrm{CR}}=\begin{cases}0 & (\tau\ge1),\\-\Lambda(1-\tau)+\kappa(1-\tau^2) & (|\tau|<1),\\-2\Lambda & (\tau\le-1);
\end{cases}
\label{eq:CR}
\end{equation}
for a partial rotation,
\begin{equation}
\Delta\omega_{\mathrm{PR}}=\begin{cases}b_1(1-\tau^2)+b_2(1-\tau^2)^2 & (|\tau|<1),\\0 & (|\tau|\ge1),
\end{cases}
\label{eq:PR}
\end{equation}
with $b_1=s_{\max}(4\kappa-2\Lambda)$ and $b_2=-4\kappa s_{\max}^2$; and for an incipient reorientation, i.e., angular wobbling about the $a$ axis,
\begin{equation}
\Delta\omega_{\mathrm{IR}}=\frac{(4\kappa-2\Lambda)\,\theta_{\mathrm{rms}}^2}{1+\tau^2},
\label{eq:IR}
\end{equation}
with $\theta_{\mathrm{rms}}$ the rms fluctuation angle at $T^*$.

The three responses are compared in Fig.~\ref{fig:theory}(b) for the same couplings. They differ in three ways. (i) Only a complete rotation leaves a step, $-2\Lambda$, below the window, because only there does the final state differ from the initial one. (ii) Both rotations are confined to $|\tau|<1$, whereas the wobbling response has Lorentzian tails, still 20\% of its maximum at $|\tau|=2$. (iii) For equal couplings the wobbling response is smaller than that of a rotation by a factor of order $\theta_{\mathrm{rms}}^2/s_{\max}$; even for $\theta_{\mathrm{rms}}=20^\circ$ it is barely visible in Fig.~\ref{fig:theory}(b). The magnetization, in turn, changes where the weak moment starts and stops turning, at the edges of a rotation, and deviates most where the rotation angle is largest, at its centre. In an incipient reorientation its anomaly lies at $T^*$, together with the phonon maximum.

$K_1$ is small near a reorientation only because its single-ion, anisotropic-exchange and Dzyaloshinskii--Moriya contributions nearly cancel [Eq.~(\ref{eq:A_K1K2})]. A phonon modulates each of them separately, whereas $\kappa$ involves only the small fourth-order anisotropy, so $|\Lambda|$ is generally expected to exceed $|\kappa|$ and a complete rotation to leave a step. Published studies agree: at complete reorientations the phonons show a step or a change of slope at the upper edge~\cite{Vilarinho2022,Mali2022,Pattanayak2021}, whereas a response confined to the window is known from the elastic constants~\cite{Belov1976,Weber2022}. A partial rotation, in contrast, leaves no step, and it is not restricted to softening: the amplitudes of Eq.~(\ref{eq:PR}) contain both couplings, $b_1$ through $4\kappa-2\Lambda$ and $b_2$ through $\kappa$ alone, and either may be positive or negative, so the same rotation can harden one mode and soften another. The ratio $b_2/b_1$ sets only the shape of the response, which can be a single bump, flatten, split into two maxima or change sign within the window [Fig.~\ref{fig:theory}(c)], while the window itself is set by the spin system and is the same for all modes. A bump without a step therefore distinguishes a partial, re-entrant rotation from a complete one, and because the window is common to all modes, whatever the sign and shape of their individual responses, it can be determined from all modes together.

\subsection{Comparison with the Raman data}
\label{sec:comparison}

\begin{figure}[H]
\centering
\includegraphics[width=\columnwidth]{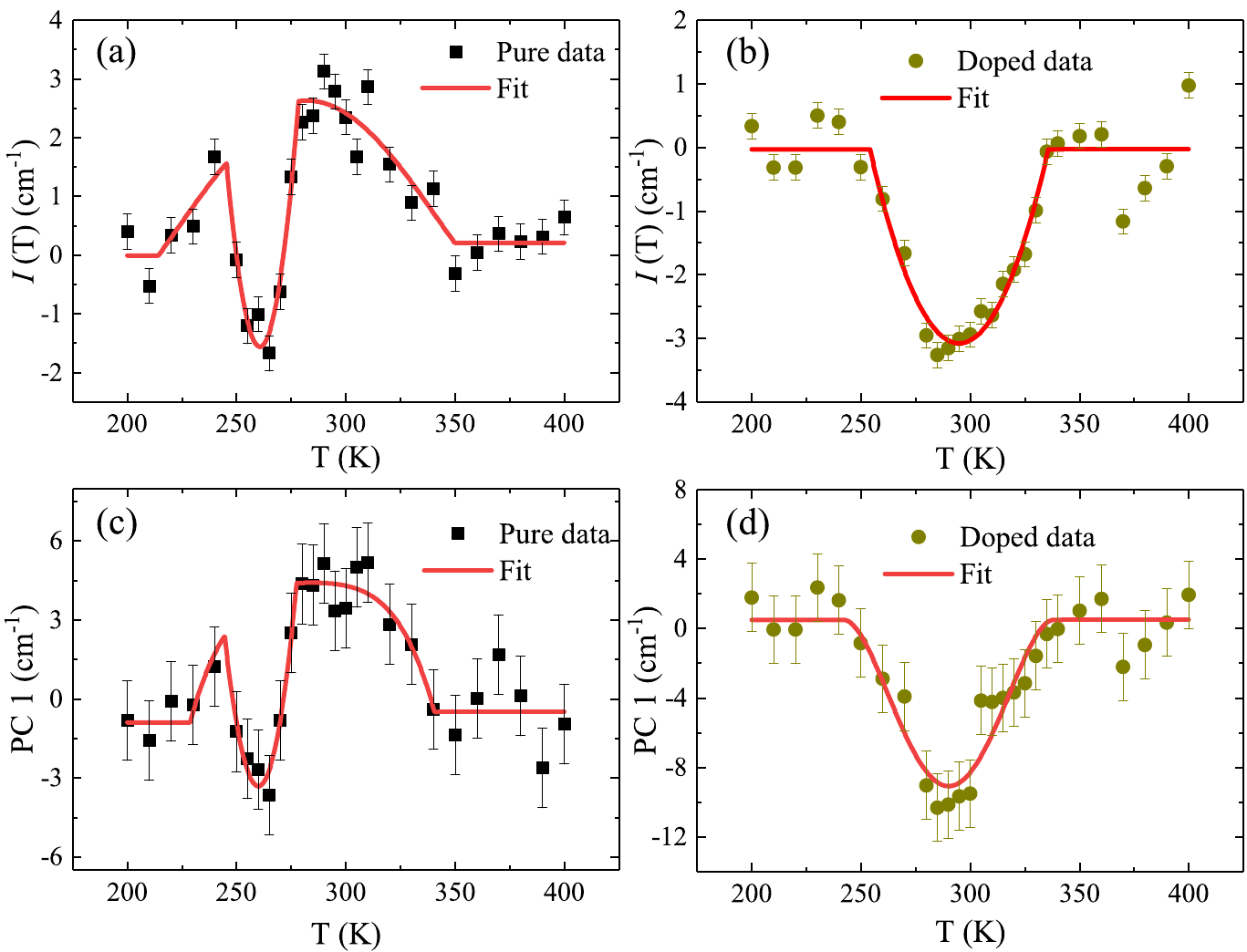}
\caption{Raman index $I(T)$ of (a) LaFeO$_3$ and (b) LaFe$_{0.9}$Ni$_{0.1}$O$_3$, and first principal component $\mathrm{PC1}(T)$ of (c) LaFeO$_3$ and (d) LaFe$_{0.9}$Ni$_{0.1}$O$_3$. Symbols are data, with error bars from the scatter outside the magnetic window; lines are fits with Eq.~(\ref{eq:PR}).}\label{fig:fits}
\end{figure}

The shifts depend on temperature only through $\langle s\rangle$ and $\langle s(1-s)\rangle$, which are common to all modes, and over our range the two terms of Eq.~(\ref{eq:PR}) are 98\% correlated. All 12 modes therefore follow one profile [Eq.~(\ref{eq:A_common})], extracted by PC1 without assuming its shape and, with the smallest uncertainty, by the Raman index $I(T)$, the noise-weighted mean of the shifts from the global fit (Appendix~\ref{app:reduction}). Figure~\ref{fig:fits} shows both fitted with Eq.~(\ref{eq:PR}), together with Eq.~(\ref{eq:CR}) for the 262~K component of LaFeO$_3$, and Table~\ref{tab:summary} lists the window centre $T_0$ and half-width $\Delta$, which agree between $I(T)$ and PC1 within two standard errors. The amplitudes $b_1$ and $b_2$ are strongly anticorrelated and are not listed.

\begin{table*}[t]
\centering
\caption{Magnetic and phonon anomalies of the two samples and parameters of the fits of Fig.~\ref{fig:fits}. $T_M$ and $\delta\chi$: temperature and size of the magnetization anomaly. $T_c$ and $\sigma$: centre and width of a Gaussian fitted to PC1, and of the global fit of the 12 modes [Eq.~(\ref{eq:A_global})]. $T_0$ and $\Delta$: centre and half-width of the partial rotation, Eq.~(\ref{eq:PR}), fitted to $I(T)$ and to PC1. For the 262~K component of LaFeO$_3$ they are $T_H$ and $\Delta_H$ of the complete rotation, Eq.~(\ref{eq:CR}). Numbers in parentheses are standard errors.}
\label{tab:summary}
\footnotesize
\setlength{\tabcolsep}{4pt}
\begin{tabular}{lccccccccc}
\hline\hline
 & \multicolumn{2}{c}{$\chi(T)$ anomaly} & \multicolumn{4}{c}{Raman anomaly} &
 \multicolumn{3}{c}{Fit with the model} \\
\cline{2-3}\cline{4-7}\cline{8-10}
 & & & \multicolumn{2}{c}{PC1, Gaussian fit} & \multicolumn{2}{c}{Global fit} & & & \\
\cline{4-5}\cline{6-7}
Sample & $T_M$ (K) & $\delta\chi$ (\%) & $T_c$ (K) & $\sigma$ (K) & $T_c$ (K) &
$\sigma$ (K) & Data & $T_0$ (K) & $\Delta$ (K) \\
\hline
\multirow{2}{*}{LaFe$_{0.9}$Ni$_{0.1}$O$_3$} & \multirow{2}{*}{323} &
\multirow{2}{*}{$-3.1$} & \multirow{2}{*}{\val{290.0(1.6)}} &
\multirow{2}{*}{\val{19.0(1.8)}} & \multirow{2}{*}{\val{290.1(1.2)}} &
\multirow{2}{*}{\val{19.0(1.4)}} & $I$ & \val{294.9(1.6)} & \val{40.8(3.2)} \\
 & & & & & & & PC1 & \val{290.2(1.9)} & \val{48.3(2.8)} \\[2pt]
\multirow{2}{*}{LaFeO$_3$} & \multirow{2}{*}{285.6} & \multirow{2}{*}{$-0.73$} &
\multirow{2}{*}{\val{290.4(5.7)}} & \multirow{2}{*}{\val{31.7(6.0)}} &
\multirow{2}{*}{\val{292.8(4.2)}} & \multirow{2}{*}{\val{33.1(5.1)}} & $I$ &
\val{282.1(4.7)} & \val{67.8(9.8)} \\
 & & & & & & & PC1 & \val{284.7(2.7)} & \val{55.9(2.9)} \\[2pt]
\multirow{2}{*}{LaFeO$_3$, 262~K} & \multirow{2}{*}{272} & \multirow{2}{*}{0.68} &
\multirow{2}{*}{\val{262.4(1.3)}} & \multirow{2}{*}{\val{8.3(1.7)}} &
\multirow{2}{*}{\val{262.2(7.9)}} & \multirow{2}{*}{\val{7.9(1.2)}} & $I$ &
\val{260.0(0.8)} & \val{16.5(1.4)} \\
 & & & & & & & PC1 & \val{261.1(1.2)} & \val{16.5(1.7)} \\
\hline\hline
\end{tabular}
\end{table*}

\begin{table*}[t]
\caption{Reorientation window, $T_0-\Delta$ to $T_0+\Delta$, and half-width $\Delta$ of the two samples [partial rotation, Eq.~(\ref{eq:PR}); ranges cover the fits to $I(T)$ and PC1] and of reported orthoferrites, with the phonon signature observed at the reorientation where available.}
\label{tab:params}
\small
\begin{tabular}{llcll}
\hline\hline
Compound & Window (K) & $\Delta$ (K) & Raman phonon signature & Ref. \\
\hline
LaFe$_{0.9}$Ni$_{0.1}$O$_3$ (partial) & \val{242--339} & \val{41--48} & bump at 290--295~K, no step & this work \\
LaFeO$_3$ (partial) & \val{214--350} & \val{56--68} &
\begin{tabular}[t]{@{}l@{}}bump at 282--285~K, no step;\\ dip at 262~K ($\alpha$-Fe$_2$O$_3$)\end{tabular} & this work \\
TmFeO$_3$ & 86--92 & 3 & & \cite{Belov1976} \\
ErFeO$_3$ & 90--100 & 5 & & \cite{Belov1976} \\
HoFeO$_3$ & 51--63 & 6 & & \cite{Belov1976} \\
SmFeO$_3$ & 470--490 & 10 & & \cite{Belov1976} \\
NdFeO$_3$ & 125--167 & 21 & step-like, from the upper edge & \cite{Belov1976,Vilarinho2022} \\
Sm$_{0.5}$Y$_{0.5}$FeO$_3$ ($\Gamma_2\to\Gamma_3$) & 282--312 & 15 & change of slope at the upper edge & \cite{Mali2022} \\
Sm$_{0.55}$Tb$_{0.45}$FeO$_3$ & 130--220 & 45 & none in the B$_{1g}$ modes tracked & \cite{Dubrovin2025} \\
TbFe$_{0.5}$Cr$_{0.5}$O$_3$ (re-entrant) & $\Gamma_4$ between 190 and 257 & & & \cite{Mali2020} \\
YFeO$_3$, EuFeO$_3$ & \begin{tabular}[t]{@{}l@{}}none (field-induced\\ at 70--75~kOe)\end{tabular} & & & \cite{Belov1976} \\
$\alpha$-Fe$_2$O$_3$ (first order) & $T_{\mathrm M}\approx250$--260 & & anomaly concentrated at $T_{\mathrm M}$ & \cite{Pattanayak2021} \\
\hline\hline
\end{tabular} 
\end{table*}

A complete rotation leaves a step below the window, which should dominate because $|\Lambda|$ is expected to exceed $|\kappa|$ (Sec.~\ref{sec:theory}). No step is observed: in both samples $I(T)$ and PC1 return to their baseline below the window (Fig.~\ref{fig:fits}). The spins thus return to $\Gamma_4$, which excludes a complete rotation. An incipient reorientation leaves no step either, but it places the phonon maximum at $T^*$, together with the magnetization anomaly. For $x=0.1$ the phonon maximum lies at 290--295~K, about 30~K below the magnetization anomaly at 323~K, which excludes this scenario. For $x=0$ the two coincide, so the position cannot decide, but the size of the response does: wobbling would need rms angles of tens of degrees [Fig.~\ref{fig:theory}(b)], and already $20^\circ$ requires thermally independent regions of only 20--40~nm [Eq.~(\ref{eq:A_volume})], incompatible with well-crystallized 400~nm grains 450~K below $T_{\mathrm N}$, where exchange keeps each grain aligned.

A partial rotation describes both samples (Fig.~\ref{fig:fits}), with windows of about \val{242}--\val{339}~K ($x=0.1$) and \val{214}--\val{350}~K ($x=0$). Within one window, modes can harden or soften, since $b_{1,i}$ and $b_{2,i}$ may have either sign [Fig.~\ref{fig:theory}(c)]; for $x=0$, B$_{2g}$(3) and A$_g$(3) harden while B$_{1g}$(3) softens. The windows also place the magnetization anomalies consistently: for $x=0.1$ the $\delta\chi$ minimum at 323~K ($\tau\approx0.7$) lies in the upper part of the window, where the spins begin to turn, and for $x=0$ it lies at the centre ($\tau\approx0$), where the angle is largest. The fourfold larger $\delta\chi$ for $x=0.1$ indicates a larger rotation.

The narrow component of LaFeO$_3$, at $T_H=\val{260}$--\val{261}~K with $\Delta_H=\val{16.5}$~K, is described by a complete rotation dominated by its $\kappa$ term, the step being about 5\% of it. It coincides with the Morin transition of trace $\alpha$-Fe$_2$O$_3$~\cite{Morin1950,Pattanayak2021}, which also produces the rise of $\chi(T)$ at 264~K.

These windows are wide (Table~\ref{tab:params}): single crystals reorient over 6--42~K~\cite{Belov1976}, and cation disorder broadens the window to about 90~K in Sm$_{0.55}$Tb$_{0.45}$FeO$_3$~\cite{Dubrovin2025}. Since the undoped sample has the wider window, the width reflects a slowly varying $K_1$ rather than Ni disorder: for $s_{\max}=\tfrac12$, $K_1$ changes by only $K_2\approx3\times10^4$~erg~cm$^{-3}$ over \val{40}--\val{70}~K, an order of magnitude more slowly than in TmFeO$_3$ ($2K_2$ over 6~K). This is expected without a magnetic rare-earth ion, whose strongly temperature-dependent coupling to Fe drives the reorientation in $R$FeO$_3$~\cite{Belov1976}.

No reorientation has been reported in LaFeO$_3$, whose Fe sublattice alone keeps $K_1>0$ in single crystals~\cite{BelovKadomtseva1971,Belov1976}. The small anomaly of our ceramic suggests that $K_1$ dips only slightly below zero, possibly through strain or defects, which alter the spin axis of LaFeO$_3$ films~\cite{Scholl2000}. Ni deepens and sharpens this dip, enlarging the rotation and narrowing the window, much as Co$^{2+}$ induces a partial, and at higher concentration a complete, reorientation in YFeO$_3$~\cite{Belov1976}.

\section{Conclusions}
\label{sec:conclusions}

We investigated the magnetic and spin-phonon responses of LaFe$_{1-x}$Ni$_x$O$_3$ ($x=0$ and 0.1) using temperature-dependent magnetization and Raman spectroscopy. Both compositions exhibit magnetic anomalies near room temperature, accompanied by correlated deviations of 12 Raman modes from their anharmonic behavior. Principal component analysis and global fitting reveal a dominant collective phonon response with mode-dependent amplitudes and signs. Comparison with a Landau-based spin-phonon coupling model favors a partial, re-entrant spin reorientation over a complete rotation or small-angle fluctuations, particularly for the Ni-substituted sample.

Within this interpretation, the N\'eel vector rotates partially from the $a$ axis toward the $c$ axis before returning to its initial orientation, with a characteristic temperature near 290~K. Ni substitution narrows the reorientation interval and enhances the magnetization anomaly by approximately a factor of four. An additional feature near 262~K in pristine LaFeO$_3$ is consistent with the Morin transition of trace $\alpha$-Fe$_2$O$_3$. These findings suggest that spin reorientation can occur in orthoferrites without magnetic rare-earth ions and that substitution at the Fe site provides a means of modifying their magnetic anisotropy. The proposed mechanism could be further examined through direct measurements of the temperature-dependent spin orientation, particularly by neutron diffraction.

\section*{Competing interests}

The authors declare no competing financial interests or personal relationships that could have influenced the work reported in this manuscript.

\section*{Acknowledgments}

S.M. and K.R.K. acknowledge financial support from the Science and Engineering Research Board (SERB), Department of Science and Technology (DST), Government of India, under Grant No.~EEQ/2022/000354. The authors thank the Central Instrumentation Facility (CIF), Indian Institute of Technology Gandhinagar, for access to characterization facilities. Financial support from DST-FIST under Grant No.~SR/FST/PSI/2020/147(C), which supported the magnetization measurement facility at IIT Gandhinagar, is also gratefully acknowledged.

\section*{Data availability}

The data supporting the findings of this study are available from the corresponding author upon reasonable request.
% =====================================================================================
% Appendix
% =====================================================================================

\appendix
\setcounter{section}{1}  
\section*{Appendix}

% -------------------------------------------------------------------------------------
\subsection{Phonon data analysis}
\label{app:reduction}

\textbf{Anharmonic baseline.} Cubic anharmonicity allows an optical phonon to decay into two phonons of about half its frequency~\cite{balkanski28}, so that
\begin{equation}
\omega_i^{\mathrm{anh}}(T)=\omega_{0,i}+C_i\left[1+\frac{2}{e^{x_i}-1}\right],
\qquad x_i=\frac{hc\,\omega_{0,i}}{2k_{\mathrm B}T},
\label{eq:A_balkanski}
\end{equation}
with the zero-temperature frequency $\omega_{0,i}$ and the anharmonic coefficient $C_i$ fitted only to temperatures outside the magnetic window: $T<250$~K and $T>310$~K for LaFeO$_3$, $T<275$~K and $T>340$~K for LaFe$_{0.9}$Ni$_{0.1}$O$_3$. These windows were fixed from $\chi(T)$ alone, so the Raman anomaly does not influence its own baseline. The anomalous shift is $\Delta\omega_i=\omega_i-\omega_i^{\mathrm{anh}}$, and its noise $\sigma_i$ is the rms of $\Delta\omega_i$ outside the window, with two degrees of freedom removed for the baseline. Only the 12 class-A modes of Table~I are used.

\textbf{Principal component analysis.} Each shift is expressed in units of its own
noise, $Z_i=\Delta\omega_i/\sigma_i$, and centred over temperature,
$\tilde Z_i=Z_i-\overline{Z_i}$. Diagonalizing the covariance matrix of the 12 modes,
\begin{equation}
C_{ij}=\frac{1}{n-1}\sum_T\tilde Z_i(T)\tilde Z_j(T),\ C\,\mathbf v_k=\lambda_k\mathbf v_k ,
\label{eq:A_pca}
\end{equation}
with $n$ the number of temperatures, gives the first principal component, $\mathrm{PC1}(T)=\sum_iv_{1,i}\tilde Z_i(T)$, with loadings $v_{1,i}$ (OriginPro: Statistics: Multivariate Analysis, covariance matrix, temperatures as observations). The sign of PC1 is chosen so that the largest loadings are positive, and its uncertainty is its rms outside the window. For 12 uncorrelated modes the largest eigenvalue exceeds 3.0 in only 5\% of simulated data sets, whereas $\lambda_1=15.5$ for LaFe$_{0.9}$Ni$_{0.1}$O$_3$ and 7.1 for LaFeO$_3$ (62\% and 44\% of the variance). PC1 thus captures the common response of the modes without assuming its shape; in LaFeO$_3$ it also contains the 262~K component.

\textbf{Global fit and Raman index.} The common profile is obtained with a smaller
uncertainty by fitting the 12 normalized shifts simultaneously,
\begin{equation}
Z_i(T)=z_{0,i}+r_i\,g(T),\qquad g(T)=\exp\!\left[-\frac{(T-T_c)^2}{2\sigma_G^2}\right],
\label{eq:A_global}
\end{equation}
with $T_c$ and $\sigma_G$ shared by all modes and an offset $z_{0,i}$ and an amplitude $r_i$ for each mode. In OriginPro this is a global fit of the 12 columns $Z_i$ with the GaussAmp function, $y_0+A\exp[-(x-x_c)^2/2w^2]$, with $x_c$ and $w$ shared and no weighting, since every $Z_i$ has unit noise; $z_{0,i}$ and $r_i$ are then $y_0$ and $A$ of curve $i$. The Gaussian only serves to determine $z_{0,i}$ and $r_i$; it belongs to none of the scenarios of Sec.~\ref{sec:theory}. For LaFe$_{0.9}$Ni$_{0.1}$O$_3$, $T_c=290.1$~K and $\sigma_G=19.2$~K. For LaFeO$_3$ the profile has two peaks, $g=G_L+c\,G_H$: a broad Gaussian $G_L$ ($T_c=292.8$~K, $\sigma_G=33.1$~K) and a narrow one, $G_H$, for the 262~K component (262.2~K, 7.9~K), with amplitudes $A_{L,i}$ and $A_{H,i}$ for each mode. Then $r_i=A_{L,i}$ and $c=\sum_iA_{L,i}A_{H,i}/\sum_iA_{L,i}^2=-0.98$.

Each mode provides its own estimate of the common profile, $e_i(T)=[Z_i(T)-z_{0,i}]/r_i$, with uncertainty $1/|r_i|$; hardening and softening modes give estimates of the same sign. Their weighted mean is the Raman index,
\begin{equation}
\begin{gathered}
I(T)=\sum_iw_i\,e_i(T)=\frac{\sum_ir_i\,[Z_i(T)-z_{0,i}]}{\sum_jr_j^2},\\
w_i=\frac{r_i^2}{\sum_jr_j^2},
\end{gathered}
\label{eq:A_index}
\end{equation}
in which each mode counts in proportion to its squared signal-to-noise ratio. Its mean outside the window is subtracted. Its uncertainty is the same at every temperature: $(\sum_jr_j^2)^{-1/2}$ for independent noise or, more conservatively, the scatter of $I$ outside the window. On average $I(T)$ follows $g(T)$, including the 262~K component of LaFeO$_3$, which enters with the weight $c$. The weights concentrate on four modes in each sample: B$_{2g}$(3), A$_g$(3), B$_{1g}$(3) and B$_{2g}$(7) in LaFeO$_3$, and A$_g$(6), B$_{2g}$(5), B$_{1g}$(4) and A$_g$(7)/B$_{3g}$ in LaFe$_{0.9}$Ni$_{0.1}$O$_3$.

\textbf{Validity and advantages of the global fit.} The global fit assumes that all modes share one temperature profile and differ only in amplitude and sign. Appendix~\ref{app:phonon} shows that this follows from the coupling of the phonons to the spin orientation [Eq.~(\ref{eq:A_common})], and the data agree: one principal component dominates in both samples, and $I(T)$ and PC1, obtained with and without a model profile, have the same shape (Fig.~\ref{fig:fits}). Compared with fits of single modes, the global fit has four advantages.
(i) \emph{Precision.} A single mode estimates the profile with noise $1/|r_i|$; $I(T)$ combines all modes with noise $(\sum_jr_j^2)^{-1/2}$, below that of the bestmode, and weak modes, whose individual fits are poorly constrained, enter only with a small weight.
(ii) \emph{Equivalence.} For fixed $z_{0,i}$ and $r_i$, and with $I$ taken before its offset is subtracted,
\begin{equation}
\begin{aligned}
&\sum_{i,T}\left[Z_i-z_{0,i}-r_i\,m(T)\right]^2\\
&\quad=\mathrm{const}+\sum_jr_j^2\sum_T\left[m(T)-I(T)\right]^2 ,
\end{aligned}
\label{eq:A_equivalence}
\end{equation}
where the constant does not depend on $m$. Fitting a model profile $m(T)$ to $I(T)$ is therefore the same as fitting $r_i\,m(T)$ to all 12 modes at once: one curve with a few parameters replaces 12 independent fits.

(iii) \emph{Robustness.} Artefacts of single modes, such as an unstable decomposition of overlapping peaks, are diluted, while the mode-specific information, the amplitudes and signs, is kept in $r_i$ and in the loadings.
The common profile is a leading-order description. In a partial rotation the ratio $b_{2,i}/b_{1,i}$, and with it the shape, may differ between modes [Eq.~(\ref{eq:A_PRi})]; such differences are of second order and are not resolved in $I(T)$ or PC1.

% -------------------------------------------------------------------------------------
\subsection{Landau free energy of the Fe spin system}
\label{app:landau}

\textbf{Order parameters.} For the four Fe moments $\mathbf M_1,\dots,\mathbf M_4$ of the orthorhombic cell, the antiferromagnetic (N\'eel) and weak-ferromagnetic vectors are
\begin{equation}
\begin{gathered}
\mathbf G=\frac{\mathbf M_1-\mathbf M_2+\mathbf M_3-\mathbf M_4}{4M_0},\\
\mathbf F=\frac{\mathbf M_1+\mathbf M_2+\mathbf M_3+\mathbf M_4}{4M_0},
\end{gathered}
\label{eq:A_GF}
\end{equation}
with $M_0$ the sublattice moment. In the canted antiferromagnetic state $|\mathbf G|\simeq1$ and $|\mathbf F|\sim10^{-2}$; $\mathbf G$ is the primary order parameter and $\mathbf F$ the weak moment generated by the Dzyaloshinskii--Moriya (DM) interaction~\cite{Dzyaloshinsky1958}.

\textbf{Reduction to $K_1$ and $K_2$.} For $\mathbf G$ in the $ac$ plane the orientation-dependent free energy is~\cite{Belov1976}
\begin{equation}
\begin{aligned}
\Phi={}&\tfrac12AF^2+\tfrac12b_1^{\mathrm a}G_x^2+\tfrac12b_3^{\mathrm a}G_z^2+\tfrac14e_1G_x^4\\
&+\tfrac14e_2G_x^2G_z^2+\tfrac14e_3G_z^4+d_1G_xF_z-d_3G_zF_x ,
\end{aligned}
\label{eq:A_Phi_full}
\end{equation}
where $A$ is the exchange stiffness against canting, $b_{1,3}^{\mathrm a}$ and $e_{1,2,3}$ are second- and fourth-order anisotropy constants (single-ion and anisotropic exchange), and $d_{1,3}$ are DM constants. Because $\mathbf F$ is small and follows $\mathbf G$, it is eliminated by
$\partial\Phi/\partial F_z=\partial\Phi/\partial F_x=0$:
\begin{equation}
\begin{gathered}
F_z=-\frac{d_1}{A}G_x,\qquad F_x=\frac{d_3}{A}G_z,\\
\tfrac12AF^2+d_1G_xF_z-d_3G_zF_x=-\frac{d_1^2G_x^2+d_3^2G_z^2}{2A}.
\end{gathered}
\label{eq:A_F}
\end{equation}
With $G_x=\cos\theta$, $G_z=\sin\theta$ and $s=\sin^2\theta$, collecting powers of $s$ gives
\begin{equation}
\Phi=\Phi_0+K_1s+K_2s^2=\Phi_0+K_1\sin^2\theta+K_2\sin^4\theta ,
\label{eq:A_Phi}
\end{equation}
\begin{equation}
\begin{gathered}
K_1=\tfrac12\big(b_3^{\mathrm a}-b_1^{\mathrm a}\big)-\tfrac12e_1+\tfrac14e_2-\frac{d_3^2-d_1^2}{2A},\\
K_2=\tfrac14(e_1-e_2+e_3).
\end{gathered}
\label{eq:A_K1K2}
\end{equation}
$K_1$ is a sum of several contributions of comparable size; it can be small, and change sign, through their near cancellation. $K_2$ involves only fourth-order terms.

\textbf{Equilibrium and stiffness.} Minimizing $\Phi$ over $0\le s\le1$, $\partial\Phi/\partial s=K_1+2K_2s=0$, gives for $K_2>0$
\begin{equation}
s=
\begin{cases}
0 & (K_1\ge0,\ \Gamma_4,\ \mathbf G\parallel a),\\
-K_1/2K_2 & (-2K_2<K_1<0,\ \Gamma_{24}),\\
1 & (K_1\le-2K_2,\ \Gamma_2,\ \mathbf G\parallel c).
\end{cases}
\label{eq:A_s}
\end{equation}
For $K_2<0$ the intermediate solution is a maximum of $\Phi$, and the spins jump between the two end states, a first-order (Morin-type) transition~\cite{Belov1976}. Using $ds/d\theta=\sin2\theta$, $\partial\Phi/\partial\theta=(K_1+2K_2s)\sin2\theta$, and the angular stiffness is
\begin{equation}
\kappa_\theta\equiv\frac{\partial^2\Phi}{\partial\theta^2}
=2(K_1+2K_2s)\cos2\theta+2K_2\sin^22\theta .
\label{eq:A_kappa_general}
\end{equation}
At equilibrium this gives $2K_1$ in $\Gamma_4$, $2K_2\sin^22\theta=8K_2s(1-s)$ in $\Gamma_{24}$ (where $K_1+2K_2s=0$) and $-2(K_1+2K_2)$ in $\Gamma_2$. The stiffness vanishes at both edges of a continuous rotation and becomes small in $\Gamma_4$ whenever $K_1\to0^+$. It also sets the frequency of the quasi-ferromagnetic magnon, $\omega_{\mathrm{qFM}}^2\propto\kappa_\theta$, whose softening at both edges of the $\Gamma_{24}$ phase, down to a magnetoelastic gap, is observed directly~\cite{Dubrovin2025}.

\textbf{Three scenarios in reduced temperature.} Near the temperature range of interest, any smooth $K_1(T)$ can be expanded about a characteristic temperature $T_0$,
\begin{equation}
K_1(T)\simeq K_1(T_0)+K_1'\,(T-T_0)+\tfrac12K_1''\,(T-T_0)^2 ,
\label{eq:A_expansion}
\end{equation}
with derivatives taken at $T_0$. We take $K_2>0$ and constant: in orthoferrites that reorient in the $ac$ plane, $K_1$ changes sign almost linearly with temperature while $K_2$ is positive and only weakly temperature dependent~\cite{Belov1976}. Which term of Eq.~(\ref{eq:A_expansion}) matters is fixed by where $T_0$ lies. At an extremum of $K_1$, $K_1'=0$ and the quadratic term leads; the sign of $K_1(T_0)$ then decides between a partial rotation and an incipient one. A complete rotation requires $K_1$ to fall from 0 to $-2K_2$ without turning back, so no extremum lies in the window and the linear term leads. Each scenario has one temperature scale $\Delta$, and in $\tau=(T-T_0)/\Delta$ every quantity depends on $\tau$ alone.

\emph{Complete rotation.} Let $T_0$ be the temperature at which $\theta=45^\circ$, i.e., $s=\tfrac12$ and $K_1(T_0)=-K_2$. Keeping the linear term, with $\alpha\equiv K_1'>0$,
\begin{equation}
\begin{gathered}
K_1(T)=-K_2+\alpha(T-T_0)=\alpha(T-T_u),\\
T_u=T_0+K_2/\alpha ,
\end{gathered}
\label{eq:A_complete_K1}
\end{equation}
the form used by Belov \textit{et al.}~\cite{Belov1976}. The rotation ends where $K_1=-2K_2$, at $T_l=T_u-2K_2/\alpha$, so that $T_0=(T_u+T_l)/2$ and $\Delta=(T_u-T_l)/2=K_2/\alpha$. With $T=T_0+\tau\Delta$,
\begin{equation}
\frac{K_1}{K_2}=\tau-1,\qquad s=\frac{1-\tau}{2}\qquad(|\tau|\le1),
\label{eq:A_complete}
\end{equation}
with $s=0$ ($\Gamma_4$) for $\tau\ge1$ and $s=1$ ($\Gamma_2$) for $\tau\le-1$: two second-order transitions in the sequence $\Gamma_4\to\Gamma_{24}\to\Gamma_2$ on cooling.

\emph{Partial (re-entrant) rotation.} If $T_0$ is a minimum of $K_1$ with $-2K_2<K_1(T_0)<0$, then $K_1=K_1(T_0)+\tfrac12K_1''(T-T_0)^2$ with $K_1''>0$, and $K_1<0$ only between its zeros at $T_0\pm\Delta$, $\Delta=\sqrt{-2K_1(T_0)/K_1''}$. Writing $K_1(T_0)=-2s_{\max}K_2$, with $0<s_{\max}<1$ the largest value of $\sin^2\theta$ reached,
\begin{equation}
\frac{K_1}{K_2}=2s_{\max}\left(\tau^2-1\right),\quad
s=s_{\max}\left(1-\tau^2\right)\quad(|\tau|\le1),
\label{eq:A_partial}
\end{equation}
and $s=0$ for $|\tau|\ge1$. The spins leave the $a$ axis at $\tau=+1$, reach $\theta_{\max}=\arcsin\sqrt{s_{\max}}$ at $\tau=0$ and return at $\tau=-1$. Re-entrant reorientations of this kind are observed in substituted orthoferrites, e.g., TbFe$_{0.5}$Cr$_{0.5}$O$_3$~\cite{Mali2020}. If $K_1(T_0)<-2K_2$ the spins reach $\Gamma_2$ and return, i.e., two complete rotations; this case is not considered.

\emph{Incipient reorientation.} If $T_0\equiv T^*$ is a minimum of $K_1$ with $K_1(T^*)>0$, $K_1$ never changes sign and the static angle stays $\theta_0=0$. The spins nevertheless fluctuate about the $a$ axis. For a small tilt $\sin^2\theta\simeq\delta\theta^2$, and the $\sin^4\theta$ term is of higher order, so a correlated region of volume $V$ has $E(\delta\theta)\simeq VK_1\delta\theta^2=\tfrac12V\kappa_\theta\delta\theta^2$ with $\kappa_\theta=2K_1$. Classical equipartition, $\langle E\rangle=\tfrac12k_{\mathrm B}T$, gives
\begin{equation}
\left\langle\delta\theta^2\right\rangle=\frac{k_{\mathrm B}T}{V\kappa_\theta}
=\frac{k_{\mathrm B}T}{2VK_1}.
\label{eq:A_equipartition}
\end{equation}
With $\Delta=\sqrt{2K_1(T^*)/K_1''}$, the distance at which $K_1$ doubles, and $k_{\min}=K_1(T^*)/K_2$,
\begin{equation}
\begin{gathered}
\frac{K_1}{K_2}=k_{\min}\left(1+\tau^2\right),\qquad
\left\langle\delta\theta^2\right\rangle=\frac{\theta_{\mathrm{rms}}^2\,T/T^*}{1+\tau^2},\\
\theta_{\mathrm{rms}}^2=\frac{k_{\mathrm B}T^*}{2VK_1(T^*)} .
\end{gathered}
\label{eq:A_incipient}
\end{equation}
The spins do not turn, but lose stiffness and fluctuate most at $\tau=0$. The small-angle form requires $\theta_{\mathrm{rms}}^2\ll k_{\min}$. Conversely, an rms angle $\theta_{\mathrm{rms}}$ at $T^*$ requires thermally independent regions of volume
\begin{equation}
V=\frac{k_{\mathrm B}T^*}{2K_1(T^*)\,\theta_{\mathrm{rms}}^2} .
\label{eq:A_volume}
\end{equation}
For $\theta_{\mathrm{rms}}=20^\circ$, $T^*=290$~K, $K_2\approx3\times10^4$~erg~cm$^{-3}$ and $K_1(T^*)=(0.1$--$1)\,K_2$, this gives $V^{1/3}\approx18$--38~nm.

% -------------------------------------------------------------------------------------
\subsection{How spin orientation shifts a phonon}
\label{app:phonon}

\textbf{Assumptions.} (i) The phonon frequency is evaluated at a fixed spin configuration: the Raman phonons (above 100~cm$^{-1}$) are much faster than the orientational dynamics of $\mathbf G$, whose lowest excitation, the quasi-ferromagnetic magnon, lies at a few cm$^{-1}$. The spins therefore do not follow the phonon displacement, and the measured shift is the frozen-spin shift averaged over the slowly varying orientations. (ii) Orientational fluctuations are described by one angle per correlated region of volume $V$. (iii) The fluctuations are small, $\langle\delta\theta^2\rangle\ll1$. (iv) They are classical: $k_{\mathrm B}T\simeq200$~cm$^{-1}$ near 290~K, much larger than the magnon energy.

\textbf{Frequency of a mode in the presence of spins.} Let $Q$ be the normal
coordinate of a Raman-active phonon with reduced mass $\mu$ and bare angular frequency
$\Omega_0$, with lattice energy $\tfrac12\mu\Omega_0^2Q^2$. The magnetic energy $E_s$
depends on $Q$ because the phonon displaces the O and La ions that set the Fe--O--Fe
bond angles and lengths. The frequency is set by the curvature of the total energy,
\begin{equation}
\mu\Omega^2=\mu\Omega_0^2+E_{s,QQ},\qquad
\Omega\simeq\Omega_0+\frac{E_{s,QQ}}{2\mu\Omega_0},
\label{eq:A_curvature}
\end{equation}
where $E_{s,QQ}\equiv\partial^2E_s/\partial Q^2$ at $Q=0$. A term linear in $Q$ only
shifts the equilibrium position (magnetostriction), and the expansion holds because
the shifts, a few cm$^{-1}$, are much smaller than the phonon frequencies. The
measured Raman shift of the Stokes line, the wavenumber of the laser minus that of the
scattered light, equals the phonon frequency in wavenumbers, $\omega=\Omega/2\pi c$.
In this unit, which is used throughout the paper,
\begin{equation}
\omega\simeq\omega_0+\frac{E_{s,QQ}}{2M},\qquad M\equiv(2\pi c)^2\mu\,\omega_0 ,
\label{eq:A_wavenumber}
\end{equation}
so that all shifts and couplings below are in cm$^{-1}$; the relative shift,
$(\omega-\omega_0)/\omega_0=E_{s,QQ}/2\mu\Omega_0^2$, is the same in either unit.
With the free energy of Appendix~\ref{app:landau},
\begin{equation}
E_s(Q,\theta)=E_{\mathrm{iso}}(Q)\,m^2+K_1(Q)\,s+K_2(Q)\,s^2 ,
\label{eq:A_Es}
\end{equation}
where $m$ is the reduced sublattice magnetization and $E_{\mathrm{iso}}m^2$ is the
isotropic exchange energy, proportional to the nearest-neighbour spin correlation
$\langle\mathbf S_i\cdot\mathbf S_j\rangle$, which is $\propto m^2$ in mean field.
Using $s=\tfrac12(1-\cos2\theta)$ and $s^2=s-\tfrac14\sin^22\theta$, and dropping
orientation-independent constants, the spin-induced shift is
\begin{equation}
\Delta\omega=\lambda^{(0)}m^2+\Lambda\cos2\theta+\kappa\sin^22\theta ,
\label{eq:A_coupling}
\end{equation}
\begin{equation}
\begin{gathered}
\lambda^{(0)}=\frac{E_{\mathrm{iso},QQ}}{2M},\\
\Lambda=-\frac{K_{1,QQ}+K_{2,QQ}}{4M},\qquad
\kappa=-\frac{K_{2,QQ}}{8M}.
\end{gathered}
\label{eq:A_couplings}
\end{equation}

\textbf{Isotropic coupling.} The term $\lambda^{(0)}m^2$ is the conventional
spin--phonon coupling~\cite{Baltensperger1968,Granado1999}. It does not depend on
$\theta$, because isotropic exchange is unchanged when all spins rotate together, and
far below $T_{\mathrm N}$ it varies smoothly with temperature. It is therefore absorbed
by the anharmonic baseline: for $T_{\mathrm N}=740$~K, a baseline fitted outside the
window reproduces the mean-field $m^2(T)$ inside it to within 5\% of its variation,
about 0.1~cm$^{-1}$ even for $\lambda^{(0)}=10$~cm$^{-1}$, below the noise of every
mode.

\textbf{Anisotropic couplings.} $\Lambda$ and $\kappa$ measure how the phonon
modulates the second- and fourth-order anisotropy; their weak dependence on $m$ is
neglected. $\Lambda$ distinguishes the two end states,
$\Delta\omega_{\Gamma_4}-\Delta\omega_{\Gamma_2}=2\Lambda$, whereas the $\kappa$ term
vanishes in both and is largest at $\theta=45^\circ$. Since a phonon modulates each of
the nearly cancelling contributions to $K_1$ [Eq.~(\ref{eq:A_K1K2})], $|\Lambda|$ is
generally expected to exceed $|\kappa|$. Averaging over small fluctuations, symmetric
about $\theta_0$, gives
\begin{equation}
\begin{aligned}
\langle\Delta\omega\rangle={}&\Lambda\cos2\theta_0\left(1-2\langle\delta\theta^2\rangle\right)\\
&+\kappa\left[\sin^22\theta_0+4\cos4\theta_0\langle\delta\theta^2\rangle\right].
\end{aligned}
\label{eq:A_fluct}
\end{equation}

\textbf{Shift in terms of $s$.} Measured from $\Gamma_4$, with $\cos2\theta-1=-2s$ and
$\sin^22\theta=4s(1-s)$, the shift of mode $i$, averaged over orientations, is
\begin{equation}
\Delta\omega_i=-2\Lambda_i\langle s\rangle+4\kappa_i\langle s(1-s)\rangle ,
\label{eq:A_shift_s}
\end{equation}
which is Eq.~(\ref{eq:shift}). The averages describe the spin system alone and are the
same for every mode. The couplings describe how the displacement pattern of mode $i$,
which involves only O and La because Fe occupies an inversion
centre~\cite{Weber2016}, modulates the anisotropy; they may have either sign, so that
some modes harden while others soften, and they are free parameters for each mode.

\textbf{The three scenarios.} Inserting $s(\tau)$ of Appendix~\ref{app:landau}, with
$\tau=(T-T_0)/\Delta$, into Eq.~(\ref{eq:A_shift_s}) gives the responses used in the
main text. The window is $|\tau|<1$; $\tau\ge1$ ($T\ge T_0+\Delta$) lies above it and
$\tau\le-1$ ($T\le T_0-\Delta$) below it. For the two rotations we use the static
orientation, $\langle s\rangle=s$ and $\langle s(1-s)\rangle=s(1-s)$. Thermal
fluctuations add corrections of order $\langle\delta\theta^2\rangle$, which round the
response near the edges of the window, where the stiffness vanishes
[Eq.~(\ref{eq:A_kappa_general})].

\emph{Complete rotation.} From Eq.~(\ref{eq:A_complete}),
\[
s=\begin{cases}
0 & (\tau\ge1),\\
(1-\tau)/2 & (|\tau|<1),\\
1 & (\tau\le-1),
\end{cases}
\]
so that inside the window $-2\Lambda_is=-\Lambda_i(1-\tau)$ and
$4\kappa_is(1-s)=\kappa_i(1-\tau^2)$, while $s(1-s)=0$ outside it. Hence
\begin{equation}
\Delta\omega_i=\begin{cases}
0 & (\tau\ge1),\\
-\Lambda_i(1-\tau)+\kappa_i(1-\tau^2) & (|\tau|<1),\\
-2\Lambda_i & (\tau\le-1),
\end{cases}
\label{eq:A_CRi}
\end{equation}
which is Eq.~(\ref{eq:CR}). The shift is continuous. It leaves zero at the upper edge
as $(2\kappa_i-\Lambda_i)(1-\tau)$, equals $\kappa_i-\Lambda_i$ at the centre, and
reaches the step at the lower edge as $-2\Lambda_i+(\Lambda_i+2\kappa_i)(1+\tau)$: the
$\Lambda$ term gives a ramp that saturates in the step $-2\Lambda_i$, and the $\kappa$
term a bump confined to the window.

\emph{Partial (re-entrant) rotation.} From Eq.~(\ref{eq:A_partial}),
\[
s=\begin{cases}
0 & (\tau\ge1),\\
s_{\max}(1-\tau^2) & (|\tau|<1),\\
0 & (\tau\le-1),
\end{cases}
\]
so that inside the window, with $P=1-\tau^2$, $-2\Lambda_is=-2\Lambda_is_{\max}P$ and
$4\kappa_is(1-s)=4\kappa_is_{\max}P-4\kappa_is_{\max}^2P^2$. Hence
\begin{equation}
\Delta\omega_i=\begin{cases}
0 & (\tau\ge1),\\
b_{1,i}(1-\tau^2)+b_{2,i}(1-\tau^2)^2 & (|\tau|<1),\\
0 & (\tau\le-1),
\end{cases}
\label{eq:A_PRi}
\end{equation}
with $b_{1,i}=s_{\max}(4\kappa_i-2\Lambda_i)$ and $b_{2,i}=-4\kappa_is_{\max}^2$; this is
Eq.~(\ref{eq:PR}). The shift is $b_{1,i}+b_{2,i}$ at the centre and rises from zero at
both edges as $2b_{1,i}(1-|\tau|)$. It is continuous, but its slope jumps at
$\tau=\pm1$, where $s$ sets in linearly, as at a second-order transition. Because the
spins return to $\Gamma_4$, the shift vanishes on both sides and there is no step. The
two amplitudes contain three unknowns, $\Lambda_i$, $\kappa_i$ and $s_{\max}$: a fit
determines $b_{1,i}$ and $b_{2,i}$, but not $s_{\max}$ separately.

\emph{Incipient reorientation.} Here $\theta_0=0$ at every temperature, and to leading
order $\langle s\rangle\simeq\langle s(1-s)\rangle\simeq\langle\delta\theta^2\rangle$
[Eq.~(\ref{eq:A_fluct})]. With Eq.~(\ref{eq:A_incipient}) and $T/T^*=1+\tau\Delta/T^*$,
\begin{equation}
\begin{aligned}
\Delta\omega_i&=(4\kappa_i-2\Lambda_i)\,\theta_{\mathrm{rms}}^2\,
\frac{1+\tau\Delta/T^*}{1+\tau^2}\\
&\simeq\frac{(4\kappa_i-2\Lambda_i)\,\theta_{\mathrm{rms}}^2}{1+\tau^2}
\qquad(\text{all }\tau),
\end{aligned}
\label{eq:A_IR}
\end{equation}
which is Eq.~(\ref{eq:IR}). The factor $T/T^*$ changes the response by a fraction
$\tau\Delta/T^*$, about $\pm14\%$ at $\tau=\pm1$ for $\Delta=40$~K and $T^*=290$~K, and
is dropped. The response has no window and no step. It is largest at the centre,
$(4\kappa_i-2\Lambda_i)\theta_{\mathrm{rms}}^2$ at $\tau=0$, falls to half of this at
$\tau=\pm1$, so that $\Delta$ is the half-width at half-maximum, and decays as
$(4\kappa_i-2\Lambda_i)\theta_{\mathrm{rms}}^2/\tau^2$ for $|\tau|\gg1$. This Lorentzian
describes $\Delta\omega$ as a function of temperature and is unrelated to the
Lorentzian component of the line shapes used to fit the individual Raman peaks.

\textbf{Common temperature profile.} In every scenario the temperature dependence
enters only through $\langle s\rangle$ and $\langle s(1-s)\rangle$, which are the
same for all modes, while the amplitudes are specific to each mode. The response of
the 12 modes therefore contains at most two independent profiles, and fewer when one
of them dominates. (i) For an incipient reorientation it contains exactly one,
$1/(1+\tau^2)$ [Eq.~(\ref{eq:A_IR})]. (ii) For a partial rotation it contains $P$ and
$P^2$ [Eq.~(\ref{eq:A_PRi})], which rise and fall together and are \val{98}\%
correlated over the measured temperatures; to leading order every mode follows one
profile with amplitude $b_{1,i}+b_{2,i}$. (iii) For a complete rotation the step and
the bump differ, and one profile describes all modes only if one term dominates in
every mode. To leading order, therefore,
\begin{equation}
\Delta\omega_i(T)\simeq y_{0,i}+A_i\,g(T),
\label{eq:A_common}
\end{equation}
with $g(T)$ common to all modes and the offset $y_{0,i}$ and amplitude $A_i$ specific
to mode $i$. In noise units, $Z_i=\Delta\omega_i/\sigma_i$, this is
Eq.~(\ref{eq:A_global}), with $z_{0,i}=y_{0,i}/\sigma_i$ and $r_i=A_i/\sigma_i$. It is
the basis of the global fit and of the Raman index, Eqs.~(\ref{eq:A_global}) and
(\ref{eq:A_index}), and PC1 recovers $g(T)$ without assuming its form
(Appendix~\ref{app:reduction}). The model predicts this common temperature
dependence and the number of its components, not which modes respond most
strongly.

\begingroup
\small
\sloppy
\bibliographystyle{apsrev4-2}
\bibliography{reference}
\endgroup

\end{multicols}
\end{document}